\documentclass[aps]{revtex4-1}

\usepackage[T1]{fontenc}
\usepackage[latin9]{inputenc}
\usepackage[a4paper]{geometry}
\usepackage{float}
\usepackage{bm}
\usepackage{units}
\usepackage{mathrsfs}
\usepackage{amsmath}
\usepackage{amssymb}
\usepackage{graphicx}
\usepackage{setspace}
\usepackage{psfrag}

\def\dbar{{\mathchar'26\mkern-12mu d}}
\newcommand{\ham}{\ensuremath{\hat{\mathscr{H}}}}

\begin{document}

\title{Statistical Thermodynamics of the Fröhlich-Bose-Einstein Condensation
of Magnons out of Equilibrium}

\author{Fabio Stucchi Vannucchi}
\affiliation{Institute of Biosciences, São Paulo State University (Unesp), São Vicente, 11330-900, SP, Brazil}
\email{fs.vannucchi@unesp.br}

\author{Áurea Rosas Vasconcellos}
\affiliation{Condensed Matter Physics Department, Institute of Physics ``Gleb Wataghin'', State University of Campinas (UNICAMP), 13083-859 Campinas, SP, Brazil}

\author{Roberto Luzzi}
\affiliation{Condensed Matter Physics Department, Institute of Physics ``Gleb Wataghin'', State University of Campinas (UNICAMP), 13083-859 Campinas, SP, Brazil}

\date{\today}

\begin{abstract}
A non-equilibrium statistical-thermodynamic approach to the study
of a Fröhlich-Bose-Einstein condensation of magnons under radio-frequency
radiation pumping is presented. Such a system displays a complex behavior
consisting in steady-state conditions to the emergence of a synergetic
dissipative structure resembling the Bose-Einstein condensation of
systems in equilibium. A kind of ``two fluid model'' arises: the
``normal'' non-equilibrium structure and Fröhlich condensate or
``non-equilibrium'' one, which is shown to be an attractor to the
system. We analyze some aspects of the irreversible thermodynamics
of this dissipative complex system, namely, its informational entropy, 
expressions for the fluctuations in non-equilibrium conditions, the 
associated Maxwell relations and the formulation of a generalized $\mathscr{H}$-theorem.
We also study the informational entropy production of the system, an 
order parameter is introduced and Glansdorff-Prigogine criteria for 
evolution and (in)stability are verified.
\end{abstract}

\maketitle

\section{Introduction}

It has been noticed \cite{kruglyak2010} that the study of collective
spin excitations in magnetically ordered materials (so-called spin
waves and the associated quasi-particles, the magnons) has a successful
history of more than 80 years \cite{bloch1930}, which recently has
re-emerged within a young field of research and technology referred-to
as \emph{Magnonics}. This term Magnonics is considered to describe
the sub-field of magnetic dynamic phenomena. The name Magnonics was
created by analogy with Electronics, with the magnons acting in the
transference of information instead of the electron charges in devices.

One important result pertaining to Magnonics has been the observation
of a macroscopic quantum phenomenon resembling a Bose-Einstein condensation
of magnons excited out of equilibrium by action of an electromagnetic
field in the radio-frequency portion of the spectrum.

The kinetic of evolution of the system of spins in thin films of yttrium-iron-garnets
(YIG) in the presence of a constant magnetic field, and being excited
by a source of rf-radiation which drives the system towards far-removed
from equilibrium conditions, has been reported in detailed experiments
performed by Demokritov et al. \cite{demokritov2006,demidov2008}.
These experimental results have evidenced the occurrence of an unexpected
large enhancement of the population of the magnons in the state lowest
in energy in their energy dispersion relation. That is, the energy
pumped on the system instead of being redistributed among the magnons
in such non-thermal conditions is transferred to the mode lowest in
frequency (with a fraction of course being dissipated to the surrounding
media). Some theoretical studies along certain approaches has been
presented by several authors (see for example Refs. \cite{tupitsyn2008,rezende2009,malomed2010});
we proceed here to describe the phenomenon within a complete thermo-statistical
description within the framework of a non-equilibrium ensemble formalism.

Such phenomenon has been referred-to as a non-equilibrium Bose-Einstein
condensation, which then would belong to a family of three types of
BEC:

The original one is the BEC in many-boson particle systems in equilibrium
at very low temperatures, which follows when their de Broglie thermal
wave length becomes larger than their mean separation distance, and
presenting some typical hallmarks (spontaneous symmetry breaking,
long-range coherence, etc.). Aside from the case of superfluidity,
BEC was realized in systems consisting of atomic alkali gases contained
in traps. A nice tutorial review is due to A. J. Leggett \cite{leggett2001}
(see also \cite{pitaevskii2003}).

A second type of BEC is the one of boson-like quasi-particles, that
is, those associated to elementary excitations in solids (e.g. phonons,
excitons, hybrid excitations, etc.), when in equilibrium at extremely
low temperatures. A well studied case is the one of an exciton-polariton
system confined in microcavities (a near two-dimensional sheet), exhibiting
the classic hallmarks of a BEC \cite{snoke2010}.

The third type, the one we are considering here, is the case of boson-like
quasi-particles (associated to elementary excitations in solids) which
are driven out of equilibrium by external perturbative sources. D.
Snoke \cite{snoke2} has properly noticed that the name BEC can be
misleading (some authors call it ``resonance'', e.g. in the case
of phonons \cite{phonon}), and following this author it is better
not to be haggling about names, and we introduce the nomenclature
NEFBEC (short for Non-Equilibrium Fröhlich-Bose-Einstein Condensation
for the reasons stated below). As noticed, here we consider the case
of magnons (boson-like quasi-particles), demonstrating that NEFBEC
of magnons is another example of a phenomenon common to many-boson
systems embedded in a thermal bath (in the conditions that the interaction
of both generates non-linear processes) when driven sufficiently away
from equilibrium by the action of an external pumping source and which
display possible applications in the technologies of devices and medicine. 
\begin{enumerate}
\item A first case was evidenced by \textbf{Herbert Fröhlich}
who considered the many boson system consisting of polar vibration
(\textsc{lo} phonons) in biopolymers under dark excitation (metabolic
energy pumping) and embedded in a surrounding fluid\cite{froehlich1970,froehlich1980,mesquita1993,fonseca2000}.
From a Science, Technology and Innovation (STI) point of view it was
considered to have implications in medical diagnosis\cite{hyland1998}.
More recently has been considered to be related to brain functioning
and artificial intelligence\cite{penrose1994}. 
\item A second case is the one of acoustic vibration (\textsc{ac}
phonons) in biological fluids, involving nonlinear anharmonic interactions
and in the presence of pumping sonic waves, with eventual STI relevance
in supersonic treatments and imaging in medicine\cite{lu1994,mesquita1998}. 
\item A third one is that of excitons (electron-hole pairs in
semiconductors) interacting with the lattice vibrations and under
the action of rf-electromagnetic fields; on a STI aspect, the phenomenon
has been considered for allowing a possible exciton-laser in the THz
frequency range called ``Excitoner''\cite{mysyrowics1996,mesquita2000}. 
\item A fourth one is the case of magnons already referred to
\cite{demokritov2006,demidov2008}, which we here analyze in depth.
The thermal bath is constituted by the phonon system, with which a
nonlinear interaction exists, and the magnons are driven arbitrarily
out of equilibrium by a source of electromagnetic radio frequency
\cite{vannucchi2010,vannucchi2013}. Technological applications are
related to the construction of sources of coherent microwave radiation
\cite{demodov2011,ma2011}. 
\end{enumerate}
There exist two other cases of NEBEC (differing from NEFBEC) where
the phenomenon is associated to the action of the pumping procedure
of drifting electron excitation, namely, 
\begin{enumerate}
\setcounter{enumi}{4}
\item A fifth one consists in a system of longitudinal acoustic
phonons driven away from equilibrium by means of drifting electron
excitation (presence of an electric field producing an electron current),
which has been related to the creation of the so-called Saser, an
acoustic laser device, with applications in computing and imaging
\cite{kent2006,rodrigues2011}. 
\item A sixth one involving a system of LO-phonons driven away
from equilibrium by means of drifting electron excitation, which displays
a condensation in an off-center small region of the Brillouin Zone
\cite{rodrigues2010,komirenko2000}. 
\end{enumerate}
We describe here item number 4, namely, a system of magnons excited
by an external pumping source. For that purpose, we consider a system
of $N$ localized spins in the presence of a constant magnetic field,
being pumped by a rf-source of radiation driving them out of equilibrium
while embedded in a thermal bath consisting of the phonon system (the
lattice vibrations) which is considered to be in equilibrium with an external reservoir at temperature
$T_{0}$. The microscopic state of the system is characterized by
the full Hamiltonian of spins and lattice vibrations after going through
Holstein-Primakov and Bogoliubov transformations\cite{keffer1966,akhiezer1968,white1983}.
On the other hand, the characterization of the macroscopic state of
the magnon system is done in terms of the Thermo-Mechanical Statistics
based on the framework of a Non-Equilibrium Statistical Ensemble Formalism
(NESEF for short)\cite{luzzilivro2002,luzzi2006,zubarev1996,kuzensky2009,akhiezer1981,mclennan1963}.
Other modern approach consists in the use of Computational Modeling\cite{kalos2007,frenkel2002}
(developed after Non-equilibrium Molecular Dynamics\cite{alder1987}).
It may be noticed that NESEF is a systematization and an extension
of the essential contributions of several renowned authors following
the brilliant pioneering work of Ludwig Boltzmann. The formalism introduces
the fundamental properties of historicity and irreversibility in the
evolution of the non-equilibrium system where dissipative and pumping
processes are under way.

In terms of the dynamics generated by the full Hamiltonian the equations
of evolution of the macroscopic state of the system are obtained in
the framework of the NESEF-based nonlinear quantum kinetic theory
\cite{luzzilivro2002,luzzi2006,zubarev1996,kuzensky2009,akhiezer1981,mclennan1963,lauck1990,kuzensky2007,madureira1998,vannucchi2009osc}.
We call the attention to the fact that the evolution equations are
the quantum mechanical equations of motion averaged over the non-equilibrium
ensemble, with the NESEF-kinetic theory providing a practical way
of calculation. The evolution of the non-equilibrium state of magnons under 
rf-radiation excitation is fully described in Refs. \cite{vannucchi2010,vannucchi2013} 
(for the sake of completeness we summarize the results in Section II),
and, on the basis of it we present here a extended study of the non-equilibrium
irreversible thermodynamics of the Fröhlich-Bose-Einstein condensation
of such ``hot'' magnons. This is done in terms of the NESEF-based Nonequilibrium-Statistical 
Irreversible Thermodynamics \cite{luzzi2000,luzzi2001} (also Ch. 7
in Ref. \cite{luzzilivro2002}).

\section{Fröhlich-Bose-Einstein Condensation of hot magnons in brief\cite{vannucchi2010,vannucchi2013}}

The system we are considering consists of a subsystem of spins being
pumped by a microwave source and interacting non-linearly with a thermal
bath (black-body radiation and crystalline lattice) that is in contact
with a thermal reservoir in equilibrium at temperature $T_{0}$. This
system is well described by the Hamiltonian 
\begin{equation}
\ham=\ham_{\mathrm{S}}+\ham_{\mathrm{Z}}+\ham_{\mathrm{SR}}+\ham_{\mathrm{R}}+\ham_{\mathrm{SL}}+\ham_{\mathrm{L}},\label{eq:hamiltoniana_total}
\end{equation}
 where $\ham_{\mathrm{S}}$ accounts for the internal (exchange and
magnetic dipole) interactions between spins, $\ham_{\mathrm{Z}}$
is associated with the effect of the constant magnetic field (Zeeman
Effect). $\ham_{\mathrm{L}}$ and $\ham_{\mathrm{R}}$ are the Hamiltonian
of the thermal bath (lattice and radiation respectively), $\ham_{\mathrm{SL}}$
and $\ham_{\mathrm{SR}}$ their interaction with the spin subsystem
($\ham_{\mathrm{SR}}$ includes also the effect of the source). Introducing
the quasi-particles related to the spin, lattice and radiation variables
(respectively the magnons, phonons and photons) and their creation
and annihilation operators ($\hat{c}_{\mathbf{q}}^{\dagger}$, $\hat{c}_{\mathbf{q}}$,
$\hat{b}_{\mathbf{q}}^{\dagger}$, $\hat{b}_{\mathbf{q}}$, $\hat{d}_{\mathbf{q}}^{\dagger}$
and $\hat{d}_{\mathbf{q}}$) we may write the Hamiltonian of Eq. (\ref{eq:hamiltoniana_total})
as 
\begin{equation}
\ham=\ham_{0}+\ham',\label{eq:separacao_hamiltoniano}
\end{equation}
 with 
\begin{align}
\ham_{0}=\; & \ham_{\mathrm{S}}^{(2)}+\ham_{\mathrm{L}}+\ham_{\mathrm{R}}=\nonumber \\
=\; & \sum_{\mathbf{q}}\hbar\omega_{\mathbf{q}}\hat{c}_{\mathbf{q}}^{\dagger}\hat{c}_{\mathbf{q}}+\sum_{\mathbf{k}}\hbar\Omega_{\mathbf{k}}\hat{b}_{\mathbf{k}}^{\dagger}\hat{b}_{\mathbf{k}}+\sum_{\mathbf{p}}\hbar\zeta_{\mathbf{p}}\hat{d}_{\mathbf{q}}^{\dagger}\hat{d}_{\mathbf{q}}\label{eq:ham_zero}
\end{align}
 being the non-interacting term formed by the Hamiltonians of free
magnons, phonons and photons, and $\hbar\omega_{\mathbf{q}}$, $\hbar\Omega_{\mathbf{k}}$
and $\hbar\zeta_{\mathbf{p}}$ their energies. The other term, 
\[
\ham'=\ham_{\mathrm{MM}}+\ham_{\mathrm{SL}}+\ham_{\mathrm{SR}},
\]
 includes the interactions between quasi-particles: 
\begin{equation}
\ham_{\mathrm{MM}}=\sum_{\mathbf{q},\mathbf{q}_{1},\mathbf{q}_{2}}\mathcal{V}_{\mathbf{q},\mathbf{q}_{1},\mathbf{q}_{2}}\hat{c}_{\mathbf{q}}^{\dagger}\hat{c}_{\mathbf{q}_{1}}^{\dagger}\hat{c}_{\mathbf{q}_{2}}\hat{c}_{\mathbf{q}+\mathbf{q}_{1}-\mathbf{q}_{2}}
\end{equation}
 is the magnon-magnon scattering term; 
\begin{align}
\ham_{\mathrm{SL}}=\; & {\displaystyle \sum_{\mathbf{q},\mathbf{k}\neq0}}(\hat{b}_{\mathbf{k}}+\hat{b}_{-\mathbf{k}}^{\dagger})\left\{ \mathcal{F}_{\mathbf{q},\mathbf{k}}\hat{c}_{\mathbf{q}}^{\dagger}\hat{c}_{\mathbf{q}-\mathbf{k}}+\mathcal{L}_{\mathbf{q},\mathbf{k}}\hat{c}_{\mathbf{q}}^{\dagger}\hat{c}_{\mathbf{k}-\mathbf{q}}^{\dagger}+\mathcal{L}_{\mathbf{q},-\mathbf{k}}^{*}\hat{c}_{\mathbf{q}}\hat{c}_{-\mathbf{k}-\mathbf{q}}\right\} +\nonumber \\
 & +{\displaystyle \sum_{\mathbf{q},\mathbf{k}\neq0}}\left\{ \mathcal{R}_{\mathbf{q},\mathbf{k}}\hat{b}_{\mathbf{k}}^{\dagger}\hat{b}_{\mathbf{k}-\mathbf{q}}+\mathcal{R}_{\mathbf{q},\mathbf{k}}^{+}\hat{b}_{\mathbf{k}}^{\dagger}\hat{b}_{\mathbf{q}-\mathbf{k}}^{\dagger}+\mathcal{R}_{-\mathbf{q},-\mathbf{k}}^{+*}\hat{b}_{-\mathbf{k}}\hat{b}_{\mathbf{k}-\mathbf{q}}\right\} (\hat{c}_{\mathbf{q}}+\hat{c}_{-\mathbf{q}}^{\dagger})
\end{align}
 accounts for the relevant magnon-phonon interaction and 
\begin{align}
\ham_{\mathrm{SR}}=\; & \sum_{\mathbf{p}}(\hat{d}_{\mathbf{p}}+\hat{d}_{-\mathbf{p}}^{\dagger})\left(\mathcal{S}_{\mathbf{p}}^{\perp*}\hat{c}_{\mathbf{p}}^{\dagger}+\mathcal{S}_{-\mathbf{p}}^{\perp}\hat{c}_{-\mathbf{p}}\right)+\nonumber \\
 & +\sum_{\mathbf{p},\mathbf{q}}(\hat{d}_{\mathbf{p}}+\hat{d}_{-\mathbf{p}}^{\dagger})\left\{ \mathcal{S}_{\mathbf{q},\mathbf{p}}^{\parallel\mathrm{a}}\hat{c}_{\mathbf{q}}^{\dagger}\hat{c}_{\mathbf{q}-\mathbf{p}}+\mathcal{S}_{\mathbf{q},\mathbf{p}}^{\parallel\mathrm{b}}\hat{c}_{\mathbf{q}}^{\dagger}\hat{c}_{\mathbf{p}-\mathbf{q}}^{\dagger}+\mathcal{S}_{\mathbf{q},-\mathbf{p}}^{\parallel\mathrm{b}*}\hat{c}_{-\mathbf{q}}\hat{c}_{\mathbf{q}-\mathbf{p}}\right\} 
\end{align}
 is the interaction between magnons and photons (source and black-body
radiation).

After the mechanical description of the system follows the thermodynamical
one. The thermodynamical state can be defined in terms of the time-dependent
thermodynamical variables 
\begin{equation}
\Biggl\{\biggl\{\mathcal{N}_{\mathbf{q}}(t)\biggr\};\:\biggl\{\mathcal{N}_{\mathbf{q},\mathbf{Q}}(t)\biggr\};\:\biggl\{\bigl\langle\hat{c}_{\mathbf{q}}^{\dagger}|t\bigr\rangle\biggr\};\:\biggl\{\bigl\langle\hat{c}_{\mathbf{q}}|t\bigr\rangle\biggr\};\:\biggl\{\sigma_{\mathbf{q}}^{\dagger}(t)\biggr\};\:\biggl\{\sigma_{\mathbf{q}}(t)\biggr\};\:\biggl\{\sigma_{\mathbf{q},\mathbf{Q}}^{\dagger}(t)\biggr\};\:\biggl\{\sigma_{\mathbf{q},\mathbf{Q}}(t)\biggr\};\: E_{\mathrm{B}}\Biggr\},\label{eq:variaveis_termodinamicas}
\end{equation}
 average values of the so-called basic micro-dynamical variables 
\begin{equation}
\Biggl\{\biggl\{\hat{\mathcal{N}}_{\mathbf{q}}\biggr\};\:\biggl\{\hat{\mathcal{N}}_{\mathbf{q},\mathbf{Q}}\biggr\};\:\biggl\{\hat{c}_{\mathbf{q}}^{\dagger}\biggr\};\:\biggl\{\hat{c}_{\mathbf{q}}\biggr\};\:\biggl\{\hat{\sigma}_{\mathbf{q}}^{\dagger}\biggr\};\:\biggl\{\hat{\sigma}_{\mathbf{q}}\biggr\};\:\biggl\{\hat{\sigma}_{\mathbf{q},\mathbf{Q}}^{\dagger}\biggr\};\:\biggl\{\hat{\sigma}_{\mathbf{q},\mathbf{Q}}\biggr\};\:\ham_{\mathrm{B}}\Biggr\},\label{eq:variaveis_de_base}
\end{equation}
with $\mathbf{Q}\neq0$, where 
\begin{equation}
\ham_{\mathrm{B}}=\ham_{\mathrm{L}}+\ham_{\mathrm{R}},
\end{equation}
 is the hamiltonian of the thermal bath (phonons and photons), 
\begin{equation}
\hat{\mathcal{N}}_{\mathbf{q}}=\hat{c}_{\mathbf{q}}^{\dagger}\hat{c}_{\mathbf{q}},\qquad\hat{\mathcal{N}}_{\mathbf{q},\mathbf{Q}}=\hat{c}_{\mathbf{q}+\frac{\mathbf{Q}}{2}}^{\dagger}\hat{c}_{\mathbf{q}-\frac{\mathbf{Q}}{2}},
\end{equation}
with $\hat{\mathcal{N}}_{\mathbf{q}}$ being the population operator
of magnons in mode $\mathbf{q}$, $\hat{\mathcal{N}}_{\mathbf{q},\mathbf{Q}}$
describing its change in space (inhomogeneities in populations) and
we recall that $\hat{c}_{\mathbf{q}}^{\dagger}$ ($\hat{c}_{\mathbf{q}}$)
are single-magnons operators whose eigenstates are the coherent states.
Finally, 
\begin{equation}
\hat{\sigma}_{\mathbf{q}}^{\dagger}=\hat{c}_{-\mathbf{q}}^{\dagger}\hat{c}_{\mathbf{q}}^{\dagger},\qquad\hat{\sigma}_{\mathbf{q},\mathbf{Q}}^{\dagger}=\hat{c}_{-\mathbf{q}-\frac{\mathbf{Q}}{2}}^{\dagger}\hat{c}_{\mathbf{q}-\frac{\mathbf{Q}}{2}}^{\dagger}
\end{equation}
 are the Hugenholtz-Gorkov pairs of two magnons.

These averages are weighted through a non-equilibrium statistical
operator $\hat{\mathscr{R}}_{\varepsilon}(t)$, for example, $\mathcal{N}_{\mathbf{q}}(t)=\mbox{Tr}\biggl\{\hat{\mathcal{N}}_{\mathbf{q}}\:\hat{\mathscr{R}}_{\varepsilon}(t)\biggr\}$.
We introduce a factorization between the thermal bath in equilibrium
and the magnetic subsystem

\begin{equation}
\hat{\mathscr{R}}_{\varepsilon}(t)=\hat{\rho}_{\varepsilon}(t)\times\hat{\rho}_{\mathrm{B}},\label{eq:neq_stat_operator}
\end{equation}
 where 

\begin{equation}
\hat{\rho}_{\mathrm{B}}=\frac{1}{Z_{\mathrm{B}}}\exp\left\{ -\beta_{\mathrm{B}}\left(\ham_{\mathrm{B}}\right)\right\} \label{eq:canonical_stat_operator}
\end{equation}
 is the canonical distribution function of the phonons and photons
in stationary condition near equilibrium at temperature $T_{\mathrm{B}}=(k_{\mathrm{B}}\beta_{\mathrm{B}})^{-1}$
(being $Z_{\mathrm{B}}$ its partition function) and $\hat{\rho}_{\varepsilon}(t)$
is the non-equilibrium statistical operator of the magnon system.
The last one may be obtained solving a modified Liouville-Dirac equation
for $\hat{\rho}_{\varepsilon}(t)$,

\begin{equation}
\frac{\partial}{\partial t}\hat{\varrho}_{\varepsilon}(t)+\frac{1}{i\hbar}\left[\hat{\varrho}_{\varepsilon}(t),\ham\right]=-\varepsilon\left\{ \hat{\varrho}_{\varepsilon}(t)-\hat{\bar{\varrho}}(t,0)\right\} ,\label{eq:liouville+fonte}
\end{equation}
 where the right term (with $\varepsilon\to0$) introduces the ``Bogoliubov's
symmetry-breaking procedure'' in time and $\hat{\bar{\varrho}}(t,0)$
is the auxiliary statistical operator. Equation (\ref{eq:liouville+fonte})
ensures on the one hand that the non-equilibrium statistical operator
$\hat{\varrho}_{\varepsilon}(t)$ incorporates the dynamical evolution
while, on the other hand, includes irreversibility \cite{luzzilivro2002,luzzi2006,zubarev1996}.

The auxiliary statistical operator $\hat{\bar{\varrho}}(t,0)$ is
written in terms of the chosen micro-dynamical variables taken the
form 
\begin{align}
\hat{\bar{\varrho}}(t,0)=\frac{1}{\bar{Z}(t)}\exp & \left\{ -{\displaystyle \sum_{\mathbf{q}}}\left[F_{\mathbf{q}}(t)\,\hat{\mathcal{N}}_{\mathbf{q}}+\phi_{\mathbf{q}}(t)\,\hat{c}_{\mathbf{q}}+\phi_{\mathbf{q}}^{*}(t)\,\hat{c}_{\mathbf{q}}^{\dagger}+\varphi_{\mathbf{q}}(t)\,\hat{\sigma}_{\mathbf{q}}+\varphi_{\mathbf{q}}^{*}(t)\,\hat{\sigma}_{\mathbf{q}}^{\dagger}\right]-\right.\nonumber \\
 & \left.-{\displaystyle \sum_{\mathbf{q},\mathbf{Q}}}\left[F_{\mathbf{q},\mathbf{Q}}(t)\,\hat{\mathcal{N}}_{\mathbf{q},\mathbf{Q}}+\varphi_{\mathbf{q},\mathbf{Q}}(t)\,\hat{\sigma}_{\mathbf{q},\mathbf{Q}}+\varphi_{\mathbf{q},\mathbf{Q}}^{*}(t)\,\hat{\sigma}_{\mathbf{q},\mathbf{Q}}^{\dagger}\right]\right\} ,\label{eq:rho_bar}
\end{align}
 where 
\begin{equation}
\Biggl\{\biggl\{ F_{\mathbf{q}}(t)\biggr\};\:\biggl\{ F_{\mathbf{q},\mathbf{Q}}(t)\biggr\};\:\biggl\{\phi_{\mathbf{q}}^{*}(t)\biggr\};\:\biggl\{\phi_{\mathbf{q}}(t)\biggr\};\:\biggl\{\varphi_{\mathbf{q}}^{*}(t)\biggr\};\:\biggl\{\varphi_{\mathbf{q}}(t)\biggr\};\:\biggl\{\varphi_{\mathbf{q},\mathbf{Q}}^{*}(t)\biggr\};\:\biggl\{\varphi_{\mathbf{q},\mathbf{Q}}(t)\biggr\}\Biggr\},\label{eq:variaveis_termodinamicas_associadas}
\end{equation}
 are the non-equilibrium thermodynamic variables conjugated to the
basic variables contained in set (\ref{eq:variaveis_termodinamicas})
in the sense of the Eqs. (\ref{eq:eq_de_estado}) and (\ref{eq:eq_de_estado_inoh})
below. The normalization of $\hat{\bar{\varrho}}(t,0)$ introduces
the non-equilibrium partition function 
\begin{align}
\bar{Z}(t)\equiv\mbox{Tr} & \left\{ \exp\left\{ -{\displaystyle \sum_{\mathbf{q}}}\left[F_{\mathbf{q}}(t)\,\hat{\mathcal{N}}_{\mathbf{q}}+\phi_{\mathbf{q}}(t)\,\hat{c}_{\mathbf{q}}+\phi_{\mathbf{q}}^{*}(t)\,\hat{c}_{\mathbf{q}}^{\dagger}+\varphi_{\mathbf{q}}(t)\,\hat{\sigma}_{\mathbf{q}}+\varphi_{\mathbf{q}}^{*}(t)\,\hat{\sigma}_{\mathbf{q}}^{\dagger}\right]-\right.\right.\nonumber \\
 & \left.\left.-{\displaystyle \sum_{\mathbf{q},\mathbf{Q}}}\left[F_{\mathbf{q},\mathbf{Q}}(t)\,\hat{\mathcal{N}}_{\mathbf{q},\mathbf{Q}}+\varphi_{\mathbf{q},\mathbf{Q}}(t)\,\hat{\sigma}_{\mathbf{q},\mathbf{Q}}+\varphi_{\mathbf{q},\mathbf{Q}}^{*}(t)\,\hat{\sigma}_{\mathbf{q},\mathbf{Q}}^{\dagger}\right]\right\} \right\} .\label{eq:particao}
\end{align}
It is important here to make three observations: first, we stress
that the auxiliary statistical operator $\hat{\bar{\varrho}}(t,0)$
does \emph{not} describe the irreversible time-evolution of the system,
and the average values weighted with $\hat{\bar{\varrho}}(t,0)$ coincide
with those weighted with $\hat{\varrho}_{\varepsilon}(t)$\emph{ only}
for the micro-dynamical variables, for example $\mathcal{N}_{\mathbf{q}}(t)=\mbox{Tr}\biggl\{\hat{\mathcal{N}}_{\mathbf{q}}\:\hat{\varrho}_{\varepsilon}(t)\times\hat{\rho}_{\mathrm{B}}\biggr\}=\mbox{Tr}\biggl\{\hat{\mathcal{N}}_{\mathbf{q}}\:\hat{\bar{\varrho}}(t,0)\times\hat{\rho}_{\mathrm{B}}\biggr\}$.
Second, the expression adopted in Eq. (\ref{eq:rho_bar}) for the
statistical operator has the form of an instantaneous generalized
canonical distribution that tends to the canonical one when the system
is in equilibrium with all the present intensive variables except $F_{\mathbf{q}}(t)$
(that, in this case, is associated with the magnons equilibrium temperature)
going to zero. Finally, since the intensive non-equilibrium thermodynamic
variables of set (\ref{eq:variaveis_termodinamicas_associadas}) equivalently
describe the macro-state of the system and that 
\begin{equation}
-\frac{\delta\ln\bar{Z}(t)}{\delta\phi_{\mathbf{q}}(t)}=\left\langle \hat{c}_{\mathbf{q}}^{\dagger}|t\right\rangle ,\quad-\frac{\delta\ln\bar{Z}(t)}{\delta F_{\mathbf{q}}(t)}=\mathcal{N}_{\mathbf{q}}(t),\quad-\frac{\delta\ln\bar{Z}(t)}{\delta\varphi_{\mathbf{q}}(t)}=\sigma_{\mathbf{q}}(t),\label{eq:eq_de_estado}
\end{equation}
\begin{equation}
-\frac{\delta\ln\bar{Z}(t)}{\delta F_{\mathbf{q},\mathbf{Q}}(t)}=\mathcal{N}_{\mathbf{q},\mathbf{Q}}(t),\quad-\frac{\delta\ln\bar{Z}(t)}{\delta\varphi_{\mathbf{q},\mathbf{Q}}(t)}=\sigma_{\mathbf{q},\mathbf{Q}}(t),\label{eq:eq_de_estado_inoh}
\end{equation}
may be considered non-equilibrium equations of state, there is a close
analogy with the intensive thermodynamic variables in equilibrium.

After presenting the relevant variables and the non-equilibrium statistical
operator, the next step in the thermodynamical description is the
derivation of the evolution equations of the thermodynamical variables
in set (\ref{eq:variaveis_termodinamicas}). Such equations form a
system of nonlinear coupled integro-differential equations which is
discussed in Ref. \cite{vannucchi2010} and in a detailed form in
Ref. \cite{vannucchi2013}. As stated there, for specific spin systems,
it suffices to follow the evolution of magnons' populations $\biggl\{\mathcal{N}_{\mathbf{q}}(t)\biggr\}$; moreover
for the equation of state it follows that 
\begin{equation}
\left\langle \hat{c}_{\mathbf{q}}^{\dagger}\hat{c}_{\mathbf{q}}|t\right\rangle =\mathcal{N}_{\mathbf{q}}(t)=\frac{1}{\mbox{e}^{F_{\mathbf{q}}(t)}-1},\label{eq:populacao(F)-1}
\end{equation}
 or, alternatively, 
\begin{equation}
F_{\mathbf{q}}(t)=\ln\left\{ 1+\frac{1}{\mathcal{N}_{\mathbf{q}}(t)}\right\} =-\ln\left\{ \frac{\mathcal{N}_{\mathbf{q}}(t)}{\mathcal{N}_{\mathbf{q}}(t)+1}\right\} .\label{eq:F(populacao)-1}
\end{equation}

We recall that the equations of evolution for the populations are
the quantum mechanical equations of motion for the dynamical quantities
$\hat{\mathcal{N}}_{\mathbf{q}}$ averaged over the non-equilibrium
ensemble. They are handled resorting to the NESEF-based nonlinear
quantum kinetic theory, with the calculations performed in the approximation
that incorporates only terms quadratic in the interaction strength
\nobreakdash- with memory and vertex renormalization neglected, that
is, we keep what in kinetic theory is called the irreducible part
of the two-particle collisions \nobreakdash- 
\begin{equation}
\frac{d}{dt}\mathcal{N}_{\mathbf{q}}(t)=\frac{1}{i\hbar}\mbox{Tr}\left\{ \left[\hat{\mathcal{N}}_{\mathbf{q}},\ham\right]\,\hat{\rho}_{\varepsilon}(t)\times\hat{\rho}_{\mathrm{B}}\right\} =J_{\mathcal{N}_{\mathbf{q}}}^{(0)}(t)+J_{\mathcal{N}_{\mathbf{q}}}^{(1)}(t)+\mathscr{J}_{\mathcal{N}_{\mathbf{q}}}^{(2)}(t),\label{eq:N_q_markov}
\end{equation}
\begin{equation}
J_{\mathcal{N}_{\mathbf{q}}}^{(0)}(t)=\frac{1}{i\hbar}\mbox{Tr}\left\{ \left[\hat{\mathcal{N}}_{\mathbf{q}},\ham_{0}\right]\,\hat{\bar{\varrho}}(t,0)\times\hat{\rho}_{\mathrm{B}}\right\} =0,\label{eq:J_0}
\end{equation}
\begin{equation}
J_{\mathcal{N}_{\mathbf{q}}}^{(1)}(t)=\frac{1}{i\hbar}\mbox{Tr}\left\{ \left[\hat{\mathcal{N}}_{\mathbf{q}},\ham'\right]\,\hat{\bar{\varrho}}(t,0)\times\hat{\rho}_{\mathrm{B}}\right\} =0,\label{eq:J_1}
\end{equation}
\begin{align}
\mathscr{J}_{\mathcal{N}_{\mathbf{q}}}^{(2)}(t)\simeq J_{\mathcal{N}_{\mathbf{q}}}^{(2)}(t)=\: & \frac{1}{(i\hbar)^{2}}\int_{-\infty}^{t}dt'\mbox{ e}^{\varepsilon(t'-t)}\mbox{ Tr}\left\{ \left[\ham'(t'-t)_{0},[\ham',\hat{\mathcal{N}}_{\mathbf{q}}]\right]\,\hat{\bar{\varrho}}(t,0)\times\hat{\rho}_{\mathrm{B}}\right\} +\nonumber \\
 & +\frac{1}{i\hbar}\sum_{\ell}\int_{-\infty}^{t}dt'\mbox{ e}^{\varepsilon(t'-t)}\mbox{Tr}\left\{ [\ham'(t'-t)_{0},\hat{P}_{\ell}]\,\hat{\bar{\varrho}}(t,0)\times\hat{\rho}_{\mathrm{B}}\right\} \frac{\delta J_{\mathcal{N}_{\mathbf{q}}}^{(1)}(t)}{\delta Q_{\ell}(t)},\label{eq:J_2}
\end{align}
 with $\hat{P}_{\ell}$ and $\hat{Q}_{\ell}$ being the variables
of sets (\ref{eq:variaveis_de_base}) and (\ref{eq:variaveis_termodinamicas})
respectively, and 
\begin{equation}
\hat{O}(t)_{0}=\mbox{e}^{-\frac{t}{i\hbar}\ham_{0}}\hat{O}\mbox{e}^{\frac{t}{i\hbar}\ham_{0}},
\end{equation}
$\delta$ stands for functional differentiation.

In a compact form we may write 
\begin{equation}
{\displaystyle \frac{d}{dt}}\mathcal{N}_{\mathbf{q}}(t)=\mathfrak{S}_{\mathbf{q}}(t)+\mathfrak{R}_{\mathbf{q}}(t)+L_{\mathbf{q}}(t)+\mathfrak{F}_{\mathbf{q}}(t)+\mathfrak{M}_{\mathbf{q}}(t),\label{eq:N_q_evol}
\end{equation}
where 
\begin{equation}
\mathfrak{S}_{\mathbf{q}}(t)=\frac{8\pi}{\hbar^{2}}\sum_{\mathbf{q}'\neq-\mathbf{q}}\left|\mathcal{S}_{\mathbf{q},\mathbf{q}+\mathbf{q}'}^{\parallel\mathrm{b}}\right|^{2}\left\{ (1+\mathcal{N}_{\mathbf{q}}+\mathcal{N}_{\mathbf{q}'})f_{\mathbf{q}'+\mathbf{q}}^{\mathrm{S}}\right\} \delta(\omega_{\mathbf{q}}+\omega_{\mathbf{q}'}-\zeta_{\mathbf{q}+\mathbf{q}'})
\end{equation}
 is the source term that accounts for the pumping of energy to the system, $f_{\mathbf{q}'+\mathbf{q}}^{\mathrm{S}}$
stands for the population of photons of the source; 
\begin{equation}
\mathfrak{R}_{\mathbf{q}}(t)={\displaystyle \frac{8\pi}{\hbar^{2}}\sum_{\mathbf{q}'\neq-\mathbf{q}}}\left|\mathcal{S}_{\mathbf{q},\mathbf{q}+\mathbf{q}'}^{\parallel\mathrm{b}}\right|^{2}\left\{ (\mathcal{N}_{\mathbf{q}'}+1)(\mathcal{N}_{\mathbf{q}}+1)f_{\mathbf{q}'+\mathbf{q}}^{\mathrm{T}}-\mathcal{N}_{\mathbf{q}'}\mathcal{N}_{\mathbf{q}}(f_{\mathbf{q}'+\mathbf{q}}^{\mathrm{T}}+1)\right\} \delta(\omega_{\mathbf{q}}+\omega_{\mathbf{q}'}-\zeta_{\mathbf{q}+\mathbf{q}'})\label{eq:R_q}
\end{equation}
 is a nonlinear term of interaction between the spin subsystem and
the black-body radiation ($f_{\mathbf{q}'+\mathbf{q}}^{\mathrm{T}}$
being its photon's population); 
\begin{equation}
L_{\mathbf{q}}(t)=-{\displaystyle \frac{1}{\tau_{\mathbf{q}}}}\left[\mathcal{N}_{\mathbf{q}}-\mathcal{N}_{\mathbf{q}}^{(0)}\right]\label{eq:L_q}
\end{equation}
is the linear relaxation to the lattice with characteristic time $\tau_{\mathbf{q}}$.
The last two terms are nonlinear contributions; 
\begin{align}
\mathfrak{F}_{\mathbf{q}}(t)=\: & {\displaystyle \frac{2\pi}{\hbar^{2}}\sum_{\mathbf{q}'\neq\mathbf{q}}}\left|\mathcal{F}_{\mathbf{q},\mathbf{q}-\mathbf{q}'}\right|^{2}\left\{ \mathcal{N}_{\mathbf{q}'}(\mathcal{N}_{\mathbf{q}}+1)(\nu_{\mathbf{q}'-\mathbf{q}}+1)-(\mathcal{N}_{\mathbf{q}'}+1)\mathcal{N}_{\mathbf{q}}\nu_{\mathbf{q}'-\mathbf{q}}\right\} \delta(\omega_{\mathbf{q}'}-\omega_{\mathbf{q}}-\Omega_{\mathbf{q}'-\mathbf{q}})+\nonumber \\
 & +{\displaystyle \frac{2\pi}{\hbar^{2}}\sum_{\mathbf{q}'\neq\mathbf{q}}}\left|\mathcal{F}_{\mathbf{q},\mathbf{q}-\mathbf{q}'}\right|^{2}\left\{ (\mathcal{N}_{\mathbf{q}}+1)\mathcal{N}_{\mathbf{q}'}\nu_{\mathbf{q}-\mathbf{q}'}-\mathcal{N}_{\mathbf{q}}(\mathcal{N}_{\mathbf{q}'}+1)(\nu_{\mathbf{q}-\mathbf{q}'}+1)\right\} \delta(\omega_{\mathbf{q}'}-\omega_{\mathbf{q}}+\Omega_{\mathbf{q}-\mathbf{q}'}),\label{eq:F_q}
\end{align}
the so-called Fröhlich term, a nonlinear interaction between magnons mediated
by the lattice, and 
\begin{equation}
\mathfrak{M}_{\mathbf{q}}(t)={\displaystyle \frac{16\pi}{\hbar^{2}}\sum_{\mathbf{q}_{1},\mathbf{q}_{2},\mathbf{q}_{3}}}\begin{array}{c}
\left|\mathcal{V}_{\mathbf{q},\mathbf{q}_{1},\mathbf{q}_{2}}\right|^{2}\left\{ (\mathcal{N}_{\mathbf{q}}+1)(\mathcal{N}_{\mathbf{q}_{1}}+1)\mathcal{N}_{\mathbf{q}_{2}}\mathcal{N}_{\mathbf{q}_{3}}-\mathcal{N}_{\mathbf{q}}\mathcal{N}_{\mathbf{q}_{1}}(\mathcal{N}_{\mathbf{q}_{2}}+1)(\mathcal{N}_{\mathbf{q}_{3}}+1)\right\} \times\\
\times\delta(\omega_{\mathbf{q}}+\omega_{\mathbf{q}_{1}}-\omega_{\mathbf{q}_{2}}-\omega_{\mathbf{q}_{3}})\delta_{\mathbf{q}_{3},\mathbf{q}+\mathbf{q}_{1}-\mathbf{q}_{2}}
\end{array},
\end{equation}
 accounts for the magnon-magnon scattering interaction term.

Although the kinetic equations for the populations {[}Eq. (\ref{eq:N_q_evol}){]}
may well describe the thermodynamic evolution of the magnetic subsystem,
the complete thermodynamic description of the entire system must also
include the evolution of the energy of the thermal bath $E_{\mathrm{B}}(t)$
(lattice and black-body radiation). In a similar form of Eq. (\ref{eq:N_q_markov})
we have that 
\begin{equation}
\frac{d}{dt}E_{\mathrm{B}}(t)=\frac{1}{i\hbar}\mbox{Tr}\left\{ \left[\ham_{\mathrm{B}},\ham\right]\,\hat{\rho}_{\varepsilon}(t)\times\hat{\rho}_{\mathrm{B}}\right\} \simeq J_{E_{\mathrm{B}}}^{(0)}(t)+J_{E_{\mathrm{B}}}^{(1)}(t)+J_{E_{\mathrm{B}}}^{(2)}(t).\label{eq:E_markov}
\end{equation}

It is simple to show that $J_{E_{\mathrm{B}}}^{(0)}(t)$ and $J_{E_{\mathrm{B}}}^{(1)}(t)$
are null. The last term is composed by two contributions: 
\begin{equation}
J_{E_{\mathrm{B}}}^{(2)}(t)=J_{\mathrm{TD}}^{(2)}(t)-\sum_{\mathbf{q}}\hbar\omega_{\mathbf{q}}\left[\mathfrak{R}_{\mathbf{q}}(t)+L_{\mathbf{q}}(t)+\mathfrak{F}_{\mathbf{q}}(t)\right],\label{eq:banho}
\end{equation}
 the first, 
\begin{equation}
J_{\mathrm{TD}}^{(2)}(t)=-\frac{E_{\mathrm{B}}(t)-E_{\mathrm{B}}^{(0)}}{\tau_{\mathrm{TD}}},
\end{equation}
 is the contribution which accounts for the thermal diffusion to the
reservoir with a thermal diffusion time $\tau_{\mathrm{TD}}$ and
tends to lead the thermal bath to equilibrium (characterized by the
equilibrium energy $E_{\mathrm{B}}^{(0)}$). The other contribution
is related to the energy received from the subsystem of magnons. 

Our system has its thermodynamical evolution described by the
kinetic equations (\ref{eq:N_q_evol}) and (\ref{eq:E_markov}) and
they must be solved. Since we stated before that the thermal bath
is in a stationary state near the equilibrium condition defined by
the external reservoir we have that 
\begin{equation}
\frac{d}{dt}E_{\mathrm{B}}(t)=J_{E_{\mathrm{B}}}^{(2)}(t)=J_{\mathrm{TD}}^{(2)}(t)-\sum_{\mathbf{q}}\hbar\omega_{\mathbf{q}}\left[\mathfrak{R}_{\mathbf{q}}(t)+L_{\mathbf{q}}(t)+\mathfrak{F}_{\mathbf{q}}(t)\right]=0,\label{eq:banho_estacionario}
\end{equation}
 and the thermal diffusion effect is sufficiently rapid for keeping
this configuration. In this case $E_{\mathrm{B}}(t)\simeq E_{\mathrm{B}}^{(0)}$,
$T_{\mathrm{B}}\simeq T_{0}$ and $\beta_{\mathrm{B}}\simeq\beta_{0}$.

Considering again the evolution of the population of magnons, we emphasise
that Eq. (\ref{eq:N_q_evol}) constitutes a nonlinear system of coupled
integro-differential equations. Its resolution in an approximate form
called ``two fluid model'' is discussed on Refs. \cite{vannucchi2010,vannucchi2013},
where the mean populations $\mathcal{N}_{1}(t)$ and $\mathcal{N}_{2}(t)$
were defined representing the populations of magnons around the minimum of
frequency and those being fed by the external source respectively,

\begin{equation}
\mathcal{N}_{1,2}(t)=\frac{\sum_{\mathbf{q}\in R_{1,2}}\mathcal{N}_{\mathbf{q}}(t)}{\sum_{\mathbf{q}\in R_{1,2}}1}=\frac{\sum_{\mathbf{q}\in R_{1,2}}\mathcal{N}_{\mathbf{q}}(t)}{n_{1,2}},\label{eq:defN1N2}
\end{equation}
 $R_{1}$ and $R_{2}$ are the correspondent regions in the reciprocal
space. Their evolution equations were obtained from Eq. (\ref{eq:N_q_evol}),

\begin{subequations}\label{eq:N1_evol-ajuste}

\begin{align}
f_{1}{\displaystyle \frac{d}{d\bar{t}}}\mathcal{N}_{1}(\bar{t})=\: & -\mathrm{D}\,\mathcal{N}_{1}(\mathcal{N}_{1}-\mathcal{N}_{1}^{(0)})-f_{1}\left[\mathcal{N}_{1}-\mathcal{N}_{1}^{(0)}\right]+\label{eq:N1_evol_relax_ajuste}\\
 & +\mathrm{F}\left\{ \mathcal{N}_{1}\mathcal{N}_{2}+\left(\bar{\nu}+1\right)\mathcal{N}_{2}-\bar{\nu}\mathcal{N}_{1}\right\} -\label{eq:N1_evol_fro_ajuste}\\
 & -\mathrm{M}\left\{ \mathcal{N}_{1}\left(\mathcal{N}_{1}+1\right)+\mathcal{N}_{2}\left(\mathcal{N}_{2}+1\right)\right\} (\mathcal{N}_{1}\frac{\mathcal{N}_{2}^{(0)}}{\mathcal{N}_{1}^{(0)}}-\mathcal{N}_{2}),\label{eq:N1_evol_mm_ajuste}
\end{align}
\end{subequations} and \begin{subequations}\label{eq:N2_evol-ajuste}
\begin{align}
f_{2}{\displaystyle \frac{d}{d\bar{t}}}\mathcal{N}_{2}(\bar{t})=\: & \mathrm{I}\,(1+2\mathcal{N}_{2})-\label{eq:N2_evol_fonte_ajuste}\\
 & -\mathrm{D}\,\mathcal{N}_{2}(\mathcal{N}_{2}-\mathcal{N}_{2}^{(0)})-f_{2}\left[\mathcal{N}_{2}-\mathcal{N}_{2}^{(0)}\right]-\label{eq:N2_evol_relax_ajuste}\\
 & -\mathrm{F}\left\{ \mathcal{N}_{1}\mathcal{N}_{2}+\left(\bar{\nu}+1\right)\mathcal{N}_{2}-\bar{\nu}\mathcal{N}_{1}\right\} +\label{eq:N2_evol_fro_ajuste}\\
 & +\mathrm{M}\left\{ \mathcal{N}_{1}\left(\mathcal{N}_{1}+1\right)+\mathcal{N}_{2}\left(\mathcal{N}_{2}+1\right)\right\} (\mathcal{N}_{1}\frac{\mathcal{N}_{2}^{(0)}}{\mathcal{N}_{1}^{(0)}}-\mathcal{N}_{2}).\label{eq:N2_evol_mm_ajuste}
\end{align}
\end{subequations} where $\bar{t}$ is the scaled time $t/\tau$,
taking the relaxation time $\tau_{\mathbf{q}}$ as having a unique
constant value ($\mathbf{q}$-independent), $\mathcal{N}_{1,2}^{(0)}$
are the populations in equilibrium, and $f_{1}$ and $f_{2}$ the
fractions of the Brillouin zone corresponding to the two regions in
the two-fluid model. Moreover, the coefficients $\mathrm{M}$ and
$\mathrm{F}$ are the coupling strengths associated to magnon-magnon
interaction and to Fröhlich contribution respectively, $\mathrm{D}$
is the one associated to decay with emission of photons, and $\bar{\nu}$
is an average population of the phonons. Finally, the parameter $\mathrm{I}$ is related to the
rate of the rf-radiation field transferred to the spin system, whose
absorption is reinforced by a positive feedback effect. All these
coefficients are dimensionless, being multiplied by the relaxation
time $\tau$. 

In a similar fashion, the energy of the thermal bath has an evolution
given, in the two fluid model, by 
\begin{align}
\frac{d}{d\bar{t}}E_{\mathrm{B}}(t)=\: & \tau J_{\mathrm{TD}}^{(2)}(t)+\nonumber \\
 & +n\hbar\omega_{1}\left\{ \mathrm{D}\,\mathcal{N}_{1}(\mathcal{N}_{1}-\mathcal{N}_{1}^{(0)})+f_{1}\left[\mathcal{N}_{1}-\mathcal{N}_{1}^{(0)}\right]-\mathrm{F}\left\{ \mathcal{N}_{1}\mathcal{N}_{2}+\left(\bar{\nu}+1\right)\mathcal{N}_{2}-\bar{\nu}\mathcal{N}_{1}\right\} \right\} +\nonumber \\
 & +n\hbar\omega_{2}\left\{ \mathrm{D}\,\mathcal{N}_{2}(\mathcal{N}_{2}-\mathcal{N}_{2}^{(0)})+f_{2}\left[\mathcal{N}_{2}-\mathcal{N}_{2}^{(0)}\right]+\mathrm{F}\left\{ \mathcal{N}_{1}\mathcal{N}_{2}+\left(\bar{\nu}+1\right)\mathcal{N}_{2}-\bar{\nu}\mathcal{N}_{1}\right\} \right\} ,\label{eq:banho_evol-ajuste}
\end{align}
being $\hbar\omega_{1}$ and $\hbar\omega_{2}$ the energy of the
magnons in the regions $R_{1}$ and $R_{2}$, and $n=\sum_{\mathbf{q}}1$.

\begin{figure}[H]
\begin{centering}
\psfrag{x}[t][c][0.85]{Time ($\mu$s)}
\psfrag{y}[b][t][0.85]{Magnon Population}
\psfrag{n1}[l][]{$\mathcal{N}_{1}$}
\psfrag{n2}[l][]{$\mathcal{N}_{2}$} 
\includegraphics[width=7cm]{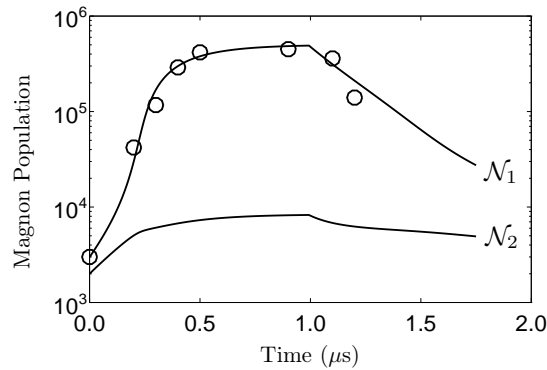}
\par\end{centering}

\caption{\label{fig:time} Evolution of the magnon population. Circles represent
Demokritov's et al. data for the low energy magnon population \cite{demokritov2006},
with the pumping being switched off after $\unit[1]{\mu s}$. Solid
lines show low and high energy magnon populations, obtained after
numerical integration of Eqs. (\ref{eq:N1_evol-ajuste}) and (\ref{eq:N2_evol-ajuste})
using the following parameters: $\mathcal{N}_{1}^{(0)}=3\times10^{3}$, $\mathcal{N}_{2}^{(0)}=2\times10^{3}$,
$f_{1}=3\times10^{-6}$, $f_{2}=3\times10^{-4}$, $\mathrm{F}=2\times10^{-6}$, $\mathrm{M}=3\times10^{-14}$, $\mathrm{D}=4\times10^{-11}$ and $\mathrm{I}=8\times10^{-4}$ . After Ref. \cite{vannucchi2010}.}
\end{figure}

On Fig. \ref{fig:time} we show the evolution of the populations $\mathcal{N}_{1}$
and $\mathcal{N}_{2}$, departing from equilibrium, under the action
of the pumping source (we adopted $\tau=\unit[1]{\mu s}$ for comparison
with experimental data \cite{demokritov2006}), solving numerically
Eqs. (\ref{eq:N1_evol-ajuste}) and (\ref{eq:N2_evol-ajuste}). As
stated on Refs. \cite{vannucchi2010,vannucchi2013}, besides the good
agreement with the experimental data, this result shows clearly
the accumulation of magnons on the mode of minimum frequency ($\mathcal{N}_{1}$).

\begin{figure}[h]
\begin{centering}
\psfrag{x}[t][c][0.85]{Scaled Rate of Pumping $\mathrm{I}$}
\psfrag{y}[b][t][0.85]{Magnon Population}
\psfrag{n1}[][][0.85]{$\mathcal{N}_{1}^{\mathrm{S}}$}
\psfrag{n2}[][][0.85]{$\mathcal{N}_{2}^{\mathrm{S}}$}
\psfrag{linear}[c][c][0.7]{\shortstack{Linear \\ Regime}}
\psfrag{bec}[c][c][0.7]{\shortstack{Emergence \\ of BEC}}
\psfrag{termico}[c][c][0.7]{\shortstack{Internal \\ thermalization}}\includegraphics[width=7cm]{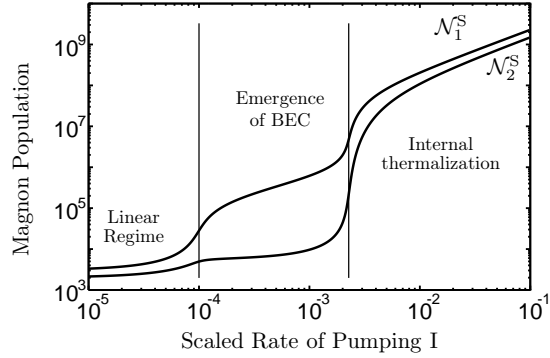}
\par\end{centering}

\caption{\label{fig:stationary}Steady-state magnon populations as a function
of the pumping source intensity which are solutions of Eqs. (\ref{eq:N1_evol-ajuste})
and (\ref{eq:N2_evol-ajuste}) using the same parameters as in Figure
\ref{fig:time}. After Ref. \cite{vannucchi2010}.}
\end{figure}

Moreover, the analysis of the steady state of the system, i. e., the
solutions of Eqs. (\ref{eq:N1_evol-ajuste}) and (\ref{eq:N2_evol-ajuste})
such that ${\displaystyle {\displaystyle \frac{d}{d\bar{t}}}\mathcal{N}_{1}(\bar{t})}$
and ${\displaystyle {\displaystyle \frac{d}{d\bar{t}}}\mathcal{N}_{2}(\bar{t})}$
are null, make evident the role of the Fröhlich term to the condensation
of magnons. On Fig. \ref{fig:stationary} we show the values of the
steady-state populations, $\mathcal{N}_{1}^{\mathrm{S}}$ and $\mathcal{N}_{2}^{\mathrm{S}}$
as a function of the scaled rate of pumping $\mathrm{I}$, and it
can be noticed the existence of two pumping scaled rate thresholds,
the first, after which there follows a steep increase in the population
of the modes lowest in frequency, corresponds to the emergence of BEC,
while the second, for higher values of $\mathrm{I}$, accounts for the internal
thermalization of the magnons which acquire a common quasi-temperature,
implying that the magnon-magnon interaction overcomes Fröhlich contribution
and BEC is impaired.

\begin{figure}[h]
\begin{centering}
\psfrag{x}[t][c][0.85]{Intensity $\mathrm{I}$}
\psfrag{y}[b][t][0.85]{Lyapunov Exponents $\lambda_\pm$}
\psfrag{l+}{$\lambda_+$}
\psfrag{l-}{$\lambda_-$}\includegraphics[width=7cm]{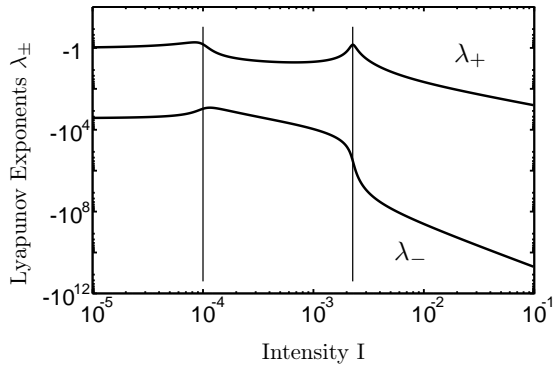}
\par\end{centering}

\caption{\label{fig:Lyapunov-Exponents}Lyapunov exponents associated with steady-state solutions of Fig. \ref{fig:stationary}.}
\end{figure}

The stability of these solutions was analyzed firstly through the
evaluation of the Lyapunov exponents. Defining the variables $a$,
$b$, $c$ and $d$ in such manner that 
\begin{align}
\frac{\partial}{\partial\mathcal{N}_{1}}\,{\displaystyle \frac{d}{d\bar{t}}}f_{1}\mathcal{N}_{1}(\bar{t})=\: & -\mathrm{D}(2\mathcal{N}_{1}^{*}-\mathcal{N}_{1}^{(0)})-f_{1}+\mathrm{F}\left\{ \mathcal{N}_{2}^{*}-\bar{\nu}\right\}+\mathrm{M}\,\left(2\mathcal{N}_{1}^{*}+1\right)\mathcal{N}_{2}^{*} -\nonumber \\
 & -\mathrm{M}\,\frac{\mathcal{N}_{2}^{(0)}}{\mathcal{N}_{1}^{(0)}}\left\{ \mathcal{N}_{1}^{*}\left(3\mathcal{N}_{1}^{*}+2\right)+\mathcal{N}_{2}^{*}\left(\mathcal{N}_{2}^{*}+1\right)\right\} \equiv a\, f_{1}\\
\frac{\partial}{\partial\mathcal{N}_{2}}\,{\displaystyle \frac{d}{d\bar{t}}}f_{1}\mathcal{N}_{1}(\bar{t})=\: & \mathrm{F}\left\{ \mathcal{N}_{1}^{*}+\bar{\nu}+1\right\} +\mathrm{M}\left\{ \mathcal{N}_{1}^{*}\left(\mathcal{N}_{1}^{*}+1\right)+\mathcal{N}_{2}^{*}\left(3\mathcal{N}_{2}^{*}+2\right)\right\}-\nonumber \\
& -\mathrm{M}\,\frac{\mathcal{N}_{2}^{(0)}}{\mathcal{N}_{1}^{(0)}}\left\{ \mathcal{N}_{1}^{*}\left(2\mathcal{N}_{2}^{*}+1\right)\right\} \equiv b\, f_{1},\\
\frac{\partial}{\partial\mathcal{N}_{1}}\,{\displaystyle \frac{d}{d\bar{t}}}f_{2}\mathcal{N}_{2}(\bar{t})=\: & -\mathrm{F}\left\{ \mathcal{N}_{2}^{*}-\bar{\nu}\right\} +\mathrm{M}\,\frac{\mathcal{N}_{2}^{(0)}}{\mathcal{N}_{1}^{(0)}}\left\{ \mathcal{N}_{1}^{*}\left(3\mathcal{N}_{1}^{*}+2\right)+\mathcal{N}_{2}^{*}\left(\mathcal{N}_{2}^{*}+1\right)\right\} -\nonumber \\
 & -\mathrm{M}\,\left(2\mathcal{N}_{1}^{*}+1\right)\mathcal{N}_{2}^{*}\equiv c\, f_{2},\\
\frac{\partial}{\partial\mathcal{N}_{2}}\,{\displaystyle \frac{d}{d\bar{t}}}f_{2}\mathcal{N}_{2}(\bar{t})=\: & 2\mathrm{I}-\mathrm{D}(2\mathcal{N}_{2}^{*}-\mathcal{N}_{2}^{(0)})-f_{2}-\mathrm{F}\left\{ \mathcal{N}_{1}^{*}+\bar{\nu}+1\right\}+ \nonumber \\
& +\mathrm{M}\,\frac{\mathcal{N}_{2}^{(0)}}{\mathcal{N}_{1}^{(0)}}\left\{ \mathcal{N}_{1}^{*}\left(2\mathcal{N}_{2}^{*}+1\right)\right\} -\nonumber \\
 & -\mathrm{M}\left\{ \mathcal{N}_{1}^{*}\left(\mathcal{N}_{1}^{*}+1\right)+\mathcal{N}_{2}^{*}\left(3\mathcal{N}_{2}^{*}+2\right)\right\}  \equiv d\, f_{2},
\end{align}
 we have that the Lyapunov exponents are the solutions of 

\begin{equation}
\left|\begin{array}{cc}
a-\lambda & b\\
c & d-\lambda
\end{array}\right|=\left(a-\lambda\right)\left(d-\lambda\right)-bc=0,
\end{equation}

 or

\begin{equation}
\lambda^{2}-\lambda\left(a+d\right)+\left(ad-bc\right)=0,
\end{equation}

 whose solutions are

\begin{equation}
\lambda_{\pm}=\frac{\left(a+d\right)}{2}\pm\sqrt{\frac{\left(a-d\right)^{2}}{4}+bc},
\end{equation}
 shown in Fig. \ref{fig:Lyapunov-Exponents} as function of $\mathrm{I}$.

Besides the expected negative values, that reflects the stability
of these solutions, it is interesting to note the two peaks occurring
precisely for the $\mathrm{I}$ values associated with the intensity
thresholds, indicating, in these regions, a possible instability of
the thermodynamic branch if we vary the values of the parameters of
Eqs. (\ref{eq:N1_evol-ajuste}) and (\ref{eq:N2_evol-ajuste}).

\section{Informational Statistical Thermodynamics of NEFBEC}

We proceed to the description of the NESEF-based Informational Irreversible
Thermodynamics, IST\cite{luzzi2000,luzzi2001}, of the NEFBEC of magnons 
\cite{demokritov2006,demidov2008} with the description given in Refs. 
\cite{vannucchi2010,vannucchi2013}, and summarized in the previous section.

\subsection{IST entropy}
\label{sec:IST_entropy}

Here, in the framework of IST, we introduce
the informational entropy 
\begin{equation}
\bar{S}(t)=-\mbox{Tr}\left\{ \hat{\mathscr{R}}_{\varepsilon}(t)\,\mathcal{P}_{\varepsilon}(t)\,\ln\hat{\mathscr{R}}_{\varepsilon}(t)\right\} ,\label{eq:entropy_formal}
\end{equation}
where, we recall, $\hat{\mathscr{R}}_{\varepsilon}(t)$ is the non-equilibrium
statistical operator of Eq. (\ref{eq:neq_stat_operator}) and $\mathcal{P}_{\varepsilon}(t)$
is a time-dependent projection operator (it is characterized by the
non-equilibrium state of the system at any time $t$) such that\cite{luzzilivro2002,luzzi1990,luzzi2000b,balian1986} 
\begin{equation}
\mathcal{P}_{\varepsilon}(t)\,\ln\hat{\rho}_{\varepsilon}(t)=\ln\hat{\bar{\varrho}}(t,0),\label{eq:projetor1}
\end{equation}
 and 
\begin{equation}
\mathcal{P}_{\varepsilon}(t)\,\ln\hat{\rho}_{\mathrm{B}}=\ln\hat{\rho}_{\mathrm{B}},\label{eq:projetor2}
\end{equation}
 where $\hat{\bar{\varrho}}(t,0)$ and $\hat{\varrho}_{\mathrm{B}}$
are those of Eqs. (\ref{eq:rho_bar}) and (\ref{eq:canonical_stat_operator})
but in the contracted description that takes as relevant micro-variables,
as stated before, only the occupation-number operator of magnons,
$\hat{\mathcal{N}}_{\mathbf{q}}$ and the Hamiltonian of the thermal
bath,$\ham_{\mathrm{B}}$, that is, in Eq. (\ref{eq:rho_bar}) the
terms involving $\phi_{\mathbf{q}}$, $\varphi_{\mathbf{q}}$, $F_{\mathbf{q},\mathbf{Q}}$,
$\hat{\sigma}_{\mathbf{q},\mathbf{Q}}$ and their conjugates are neglected.

Hence we have that 
\begin{equation}
\bar{S}(t)=-\mbox{Tr}\left\{ \hat{\mathscr{R}}_{\varepsilon}(t)\,\ln\left\{ \hat{\bar{\varrho}}(t,0)\times\hat{\rho}_{\mathrm{B}}\right\} \right\} =\phi(t)+\beta_{\mathrm{B}}E_{\mathrm{B}}+{\displaystyle \sum_{\mathbf{q}}}F_{\mathbf{q}}(t)\,\mathcal{N}_{\mathbf{q}}(t),\label{eq:ie_bosons_1}
\end{equation}

\begin{equation}
\phi(t)=\ln Z_{\mathrm{B}}+\ln\bar{Z}(t),\label{generating_functional}
\end{equation}
where $Z_{\mathrm{B}}$ and $\bar{Z}(t)$ are the canonical and non-equilibrium
partition functions {[}see Eqs. (\ref{eq:canonical_stat_operator})
and (\ref{eq:particao}){]}. The last one depends on time and must
be explicitly written in terms of the non-equilibrium thermodynamic
variables. In a analogous way to the equilibrium Bose-statistics we
obtain that 
\begin{align}
\bar{Z}(t)=\: & \mbox{Tr}\exp\{-{\displaystyle \sum_{\mathbf{q}}}F_{\mathbf{q}}(t)\,\hat{\mathcal{N}}_{\mathbf{q}}\}=\prod_{\mathbf{q}}\frac{1}{1-\mbox{e}^{-F_{\mathbf{q}}(t)}}.
\end{align}
 Thus 
\begin{equation}
\ln\bar{Z}(t)={\displaystyle \sum_{\mathbf{q}}}\ln\frac{1}{1-\mbox{e}^{-F_{\mathbf{q}}(t)}}
\end{equation}
 and using the relation between $F_{\mathbf{q}}(t)$ and $\mathcal{N}_{\mathbf{q}}(t)$,
Eq. (\ref{eq:F(populacao)-1}), one obtains the expression for the
informational entropy 
\begin{equation}
\bar{S}(t)=\ln Z_{\mathrm{B}}+\beta_{\mathrm{B}}E_{\mathrm{B}}-{\displaystyle \sum_{\mathbf{q}}}\left\{ \mathcal{N}_{\mathbf{q}}(t)\,\ln\left[\mathcal{N}_{\mathbf{q}}(t)\right]-\left[\mathcal{N}_{\mathbf{q}}(t)+1\right]\,\ln\left[\mathcal{N}_{\mathbf{q}}(t)+1\right]\right\} ,\label{eq:ie_bosons_2}
\end{equation}
 whereas $Z_{\mathrm{B}}$, $\beta_{\mathrm{B}}$ and $E_{\mathrm{B}}$
are constants {[}see Eq. (\ref{eq:banho_estacionario}) and subsequent
discussion{]}.

\subsection{Fluctuations and Maxwell Relations}
\label{sec:fluctuations}

As already shown, the average value of any dynamical quantity of the
basic set in NESEF is given by minus the functional derivative of
the generating functional $\phi(t)$ with respect to the associated
non-equilibrium thermodynamic variables {[}and we recall that this
function can be related to a kind of non-equilibrium partition function
through the expression $\phi(t)=\ln Z_{\mathrm{B}}+\ln\bar{Z}(t)$,
cf. Eqs. (\ref{eq:particao}), (\ref{eq:eq_de_estado}), (\ref{eq:eq_de_estado_inoh})
and (\ref{generating_functional}){]}. Considering only the
populations of magnons as relevant variables, 
\begin{equation}
\bar{Z}(t)=\mbox{Tr}\,\exp\left\{ -{\displaystyle \sum_{\mathbf{q}}}F_{\mathbf{q}}(t)\,\hat{\mathcal{N}}_{\mathbf{q}}\right\} ,
\end{equation}
 ignoring here and in what follows the constant $Z_{\mathrm{B}}$, we have that 
\begin{align}
-\frac{\delta\phi(t)}{\delta F_{\mathbf{q}}(t)}=\: & -\frac{1}{\bar{Z}(t)}\frac{\delta\bar{Z}(t)}{\delta F_{\mathbf{q}}(t)}=-\frac{1}{\bar{Z}(t)}\mbox{Tr}\,\frac{\delta}{\delta F_{\mathbf{q}}(t)}\exp\left\{ -{\displaystyle \sum_{\mathbf{q}}}F_{\mathbf{q}}(t)\,\hat{\mathcal{N}}_{\mathbf{q}}\right\} =\nonumber \\
=\: & \frac{1}{\bar{Z}(t)}\mbox{Tr}\,\left\{\hat{\mathcal{N}}_{\mathbf{q}}\,\exp\left\{ -{\displaystyle \sum_{\mathbf{q}}}F_{\mathbf{q}}(t)\,\hat{\mathcal{N}}_{\mathbf{q}}\right\}\right\} =\mathcal{N}_{\mathbf{q}}(t).
\end{align}

Moreover, from a straight calculation it follows that 
\begin{equation}
\frac{\delta^{2}\phi(t)}{\delta F_{\mathbf{q}'}(t)\,\delta F_{\mathbf{q}''}(t)}=-\frac{\delta\mathcal{N}_{\mathbf{q}'}(t)}{\delta F_{\mathbf{q}''}(t)}=-\frac{\delta\mathcal{N}_{\mathbf{q}''}(t)}{\delta F_{\mathbf{q}'}(t)}=\mbox{Tr}\left\{ \Delta\hat{\mathcal{N}}_{\mathbf{q}'}\Delta\hat{\mathcal{N}}_{\mathbf{q}''}\,\hat{\bar{\varrho}}(t,0)\right\} =\mathcal{C}_{\mathbf{q}'\mathbf{q}''}(t),\label{eq:correlations}
\end{equation}
 where 
\begin{equation}
\Delta\hat{\mathcal{N}}_{\mathbf{q}'}=\hat{\mathcal{N}}_{\mathbf{q}'}-\mbox{Tr}\left\{ \hat{\mathcal{N}}_{\mathbf{q}'}\,\hat{\bar{\varrho}}(t,0)\right\} =\hat{\mathcal{N}}_{\mathbf{q}'}-\mathcal{N}_{\mathbf{q}'}(t),
\end{equation}
 and Eq. (\ref{eq:correlations}) defines the \emph{matrix of correlations}
$\hat{\mathcal{C}}(t)$. Their diagonal elements are the mean square
deviations, or fluctuations, of quantities $\hat{\mathcal{N}}_{\mathbf{q}'}$,
namely 
\begin{equation}
\mathcal{C}_{\mathbf{q}\mathbf{q}}(t)=\mbox{Tr}\left\{ \left[\Delta\hat{\mathcal{N}}_{\mathbf{q}}\right]^{2}\,\hat{\bar{\varrho}}(t,0)\right\} =\mbox{Tr}\left\{ \left[\hat{\mathcal{N}}_{\mathbf{q}}-\mathcal{N}_{\mathbf{q}}(t)\right]^{2}\,\hat{\bar{\varrho}}(t,0)\right\} \equiv\Delta^{2}\mathcal{N}_{\mathbf{q}}(t)
\end{equation}
 and the matrix is symmetrical, that is, 
\begin{equation}
\mathcal{C}_{\mathbf{q}'\mathbf{q}''}(t)=\frac{\delta^{2}\phi(t)}{\delta F_{\mathbf{q}'}(t)\,\delta F_{\mathbf{q}''}(t)}=\frac{\delta^{2}\phi(t)}{\delta F_{\mathbf{q}''}(t)\,\delta F_{\mathbf{q}'}(t)}=\mathcal{C}_{\mathbf{q}''\mathbf{q}'}(t)
\end{equation}
 what is a manifestation in IST of the known Maxwell relations in
equilibrium.

Let us next scale the informational entropy and the non-equilibrium
thermodynamic intensive variables in terms of Boltzmann constant,
$k_{\mathrm{B}}$, that is, we introduce 
\begin{equation}
\bar{\mathcal{S}}(t)=k_{\mathrm{B}}\bar{S}(t);\qquad\mathbb{F}_{\mathbf{q}}(t)=k_{\mathrm{B}}F_{\mathbf{q}}(t);
\end{equation}
 and then, because of Eq. (\ref{eq:ie_bosons_1}), 
\begin{equation}
\mathbb{F}_{\mathbf{q}}(t)=\frac{\delta\bar{\mathcal{S}}(t)}{\delta\mathcal{N}_{\mathbf{q}}(t)}.\label{eq:associate}
\end{equation}

Moreover, we find that 
\begin{equation}
\frac{\delta^{2}\bar{\mathcal{S}}(t)}{\delta\mathcal{N}_{\mathbf{q}'}(t)\,\delta\mathcal{N}_{\mathbf{q}''}(t)}=\frac{\delta\mathbb{F}_{\mathbf{q}'}(t)}{\delta\mathcal{N}_{\mathbf{q}''}(t)}=\frac{\delta\mathbb{F}_{\mathbf{q}''}(t)}{\delta\mathcal{N}_{\mathbf{q}'}(t)}=-k_{\mathrm{B}}\mathcal{C}_{\mathbf{q}'\mathbf{q}''}^{(-1)}(t),
\end{equation}
that is, the second order functional derivatives of the IST-informational-entropy
are the components of minus the inverse of the matrix of correlations
$\mathcal{C}^{(-1)}$, with elements to be denoted by $\mathcal{C}_{\mathbf{q}'\mathbf{q}''}^{(-1)}$.
Besides, the fluctuation of the IST-informational-entropy is given
by 
\begin{equation}
\Delta^{2}\bar{\mathcal{S}}(t)=\sum_{\mathbf{q}'\,\mathbf{q}''}\frac{\delta\bar{\mathcal{S}}(t)}{\delta\mathcal{N}_{\mathbf{q}'}(t)}\frac{\delta\bar{\mathcal{S}}(t)}{\delta\mathcal{N}_{\mathbf{q}''}(t)}\mathcal{C}_{\mathbf{q}'\mathbf{q}''}(t)=\sum_{\mathbf{q}'\,\mathbf{q}''}\mathcal{C}_{\mathbf{q}'\mathbf{q}''}(t)\,\mathbb{F}_{\mathbf{q}'}(t)\,\mathbb{F}_{\mathbf{q}''}(t),
\end{equation}
 and that of the non-equilibrium thermodynamic variables $\mathbb{F}_{\mathbf{q}}(t)$
are 
\begin{equation}
\Delta^{2}\mathbb{F}_{\mathbf{q}}(t)=\sum_{\mathbf{q}'\,\mathbf{q}''}\frac{\delta\mathbb{F}_{\mathbf{q}}(t)}{\delta\mathcal{N}_{\mathbf{q}'}(t)}\frac{\delta\mathbb{F}_{\mathbf{q}}(t)}{\delta\mathcal{N}_{\mathbf{q}''}(t)}\mathcal{C}_{\mathbf{q}'\mathbf{q}''}(t)=k_{\mathrm{B}}^{2}\sum_{\mathbf{q}'\,\mathbf{q}''}\mathcal{C}_{\mathbf{q}\mathbf{q}'}^{(-1)}(t)\,\mathcal{C}_{\mathbf{q}\mathbf{q}''}^{(-1)}(t)\,\mathcal{C}_{\mathbf{q}'\mathbf{q}''}(t)=k_{\mathrm{B}}^{2}\mathcal{C}_{\mathbf{q}\mathbf{q}}^{(-1)}(t),
\end{equation}
therefore 
\begin{equation}
\Delta^{2}\mathcal{N}_{\mathbf{q}}(t)\,\Delta^{2}\mathbb{F}_{\mathbf{q}}(t)=k_{\mathrm{B}}^{2}\mathcal{C}_{\mathbf{q}\mathbf{q}}(t)\,\mathcal{C}_{\mathbf{q}\mathbf{q}}^{(-1)}(t)=k_{\mathrm{B}}^{2}G_{\mathbf{q}\mathbf{q}}(t),
\end{equation}
 where 
\begin{equation}
G_{\mathbf{q}\mathbf{q}}(t)=\mathcal{C}_{\mathbf{q}\mathbf{q}}(t)\,\mathcal{C}_{\mathbf{q}\mathbf{q}}^{(-1)}(t),
\end{equation}
 and then 
\begin{equation}
\left[\Delta^{2}\mathcal{N}_{\mathbf{q}}(t)\right]^{\nicefrac{1}{2}}\,\left[\Delta^{2}\mathbb{F}_{\mathbf{q}}(t)\right]^{\nicefrac{1}{2}}=k_{\mathrm{B}}\left[G_{\mathbf{q}\mathbf{q}}(t)\right]^{\nicefrac{1}{2}}.\label{eq:uncertainty}
\end{equation}

The quantities $\mathcal{C}_{\mathbf{q}'\mathbf{q}''}^{(-1)}$ are
the matrix elements of the inverse of the matrix of correlations,
and if the variables are uncorrelated $G_{\mathbf{q}\mathbf{q}}(t)=1$.
Equation (\ref{eq:uncertainty}) has the likeness of an uncertainty
principle connecting the variables $\mathcal{N}_{\mathbf{q}}(t)$
and $\mathbb{F}_{\mathbf{q}}(t)$, which are thermodynamically conjugated
in the sense of Eqs. (\ref{eq:eq_de_estado}) and (\ref{eq:associate}),
with Boltzmann constant being the atomistic parameter playing a role
resembling that of the quantum of action in mechanics. This leads
to the possibility to relate the results of IST with the idea of complementarity
between the microscopic and macroscopic descriptions of many-body
systems advanced by Rosenfeld and Prigogine \cite{rosenfeld1979,rosenfeld1960,rosenfeld1955,prigogine1980};
this is discussed elsewhere \cite{luzzi1998}.

Care must be exercised in referring to fluctuations of the intensive
variables $F_{\mathbf{q}}$. In the statistical description fluctuations
are associated to the specific variables $\mathcal{N}_{\mathbf{q}}$,
but the $F_{\mathbf{q}}$ are non-equilibrium thermodynamic intensive
variables fixed by the average values of the $\hat{\mathcal{N}}_{\mathbf{q}}$,
and so $\Delta^{2}F_{\mathbf{q}}$ is not a proper fluctuation of
$F_{\mathbf{q}}$ but a second order deviation interpreted as being
a result of the fluctuations of the variables on which it depends,
in a generalization of the usual results in statistical mechanics
in equilibrium \cite{kittel1988}. These brief considerations point
to the desirability to develop a complete theory of fluctuations in
the context of NESEF; one relevant application of it would be the
study of the kinetics of transition between dissipative structures
in complex systems, of which is presently available a phenomenological
approach \cite{nicolis1977}.

\subsection{A Boltzmann-like relation: \textmd{\normalsize{}$\bar{\mathcal{S}}(t)=k_{\mathrm{B}}\ln W(t)$}}
\label{sec:boltzmann}

According to the results of the previous subsection, quite similarly
to the case of equilibrium it follows that the quotient between the
root mean square of a given quantity and its average value is of the
order of the reciprocal of the square root of the number of particles,
that is 
\begin{equation}
\frac{\left[\Delta^{2}\mathcal{N}_{\mathbf{q}}(t)\right]^{\nicefrac{1}{2}}}{\mathcal{N}_{\mathbf{q}}(t)}\sim N^{-\nicefrac{1}{2}}
\end{equation}

Consequently, again quite in analogy with the case of equilibrium,
the number of states contributing for the quantity $N$ to have the
given average value, is overwhelmingly enormous (a rigorous demonstration
follows resorting to the method of the steepest descent \cite{pathria1972}).
Therefore, we can write that 
\begin{equation}
\phi(t)=\ln\,\mbox{Tr}\,\exp\left\{ -{\displaystyle \sum_{\mathbf{q}}}F_{\mathbf{q}}(t)\,\hat{\mathcal{N}}_{\mathbf{q}}\right\} \simeq\ln\left\{ W(t)\,\exp\left\{ -{\displaystyle \sum_{\mathbf{q}}}F_{\mathbf{q}}(t)\,\hat{\mathcal{N}}_{\mathbf{q}}\right\} \right\} ,\label{eq:lnW}
\end{equation}
 where 
\begin{equation}
W(t)=\sum_{\tilde{n}\in\mathcal{M}(t)}1=\mbox{number of states in }\mathcal{M}(t),
\end{equation}
 where $\tilde{n}$ is the set of quantum numbers which characterize
the quantum-mechanical state of the system, and $\mathcal{M}$ contains
the set of states $\left|\tilde{n}\right\rangle $ such that 
\begin{equation}
\mathcal{M}(t)\,:\,\mathcal{N}_{\mathbf{q}}(t)\leq\left\langle \tilde{n}\right|\hat{\mathcal{N}}_{\mathbf{q}}\left|\tilde{n}\right\rangle \leq\mathcal{N}_{\mathbf{q}}(t)+\Delta\mathcal{N}_{\mathbf{q}}(t),
\end{equation}
 where we have used the usual notations of bra and ket and matrix
elements between those states. Hence we have that {[}cf. Eq. (\ref{eq:ie_bosons_1}){]} 
\begin{equation}
\bar{\mathcal{S}}(t)=k_{\mathrm{B}}\bar{S}(t)=k_{\mathrm{B}}\phi(t)+k_{\mathrm{B}}{\displaystyle \sum_{\mathbf{q}}}F_{\mathbf{q}}(t)\,\mathcal{N}_{\mathbf{q}}(t)\simeq k_{\mathrm{B}}\ln W(t),\label{eq:boltzmann}
\end{equation}
 using Eq. (\ref{eq:lnW}), after disregarding the constant contribution
from the thermal bath. 

We recall that this is an approximate result, with an error of the
order of the reciprocal of the square root of the number of degrees
of freedom of the system, and therefore exact only in the thermodynamic
limit.

Equation (\ref{eq:boltzmann}) represents the equivalent of Boltzmann
expression for the thermodynamic entropy in terms of the logarithm
of the number of complexions compatible with the macroscopic constraints
imposed on the system. 

Citing Jaynes, it is this property of the entropy - measuring our
degree of information about the microstate, which is conveyed by data
on the macroscopic thermodynamic variables - that made information
theory such a powerful tool in showing us how to generalize Gibbs'
equilibrium ensembles to non-equilibrium ones. The generalization
could never have been found by those who thought that entropy was,
like energy, a physical property of the microstate \cite{jaynes1988}.
Also following Jaynes, $W(t)$ measures the degree of control of the
experimenter over the microstate, when the only parameters the experimenter
can manipulate are the usual macroscopic ones. At time $t$, when
a measurement is performed, the state is characterized by the set
$\left\{ \mathcal{N}_{\mathbf{q}}(t)\right\} $, and the corresponding
phase volume is $W(t)$, containing all conceivable ways in which
the final macrostate can be realized. But, since the experiment is
to be reproducible, the region with volume $W(t)$ should contain
at least the phase points originating in the region of volume $W(t_{0})$,
and then $W(t)>W(t_{0})$. Because phase volume is conserved in the
micro-dynamical evolution, it is a fundamental requirement on any
reproducible process that the phase volume $W(t)$ compatible with
the final state cannot be less than the phase volume $W(t_{0})$ which
describes our ability to reproduce the initial state \cite{jaynes1965}.

\subsection{IST entropy and order parameter for magnons}
\label{sec:order_parameter}

In the cited ``two-fluid model'' the informational entropy, neglecting
the constant part related to the bath, may be written as 
\begin{align}
\bar{S}(t)=-n_{1}\left\{ \mathcal{N}_{1}\,\ln\left(\mathcal{N}_{1}\right)-\left(\mathcal{N}_{1}+1\right)\,\ln\left(\mathcal{N}_{1}+1\right)\right\} - \nonumber \\ 
- n_{2}\left\{ \mathcal{N}_{2}\,\ln\left(\mathcal{N}_{2}\right)-\left(\mathcal{N}_{2}+1\right)\,\ln\left(\mathcal{N}_{2}+1\right)\right\} ,\label{eq:ie_2_fluidos}
\end{align}
 omitting to indicate the time dependence on the right for practical
convenience. 

The informational entropy in IST also satisfies a kind of generalized
Clausius relation. In fact, consider the modification of the informational
entropy as a consequence of the modification of external constraints
imposed on the system. Let us call $\lambda_{\ell}(\ell=1,2,\ldots,s)$
a set of parameters that characterize such constraints (e.g., the
volume, external fields, etc.). Introducing infinitesimal modifications
of them, say $d\lambda_{\ell}$, the corresponding variation in the
informational entropy, in the two-fluid model that was introduced, is given by 
\begin{equation}
d\bar{S}(t)=F_{1}(t)\:\dbar\mathcal{N}_{1}(t)+F_{2}(t)\:\dbar\mathcal{N}_{2}(t)\label{eq:differential}
\end{equation}
 where $\dbar\mathcal{N}_{1,2}(t)$ are the nonexact differentials
\begin{equation}
\dbar\mathcal{N}_{1,2}(t)=d\mathcal{N}_{1,2}(t)-\left\langle d\hat{\mathcal{N}}_{1,2}|t\right\rangle ,
\end{equation}
 with $\left\langle d\hat{\mathcal{N}}_{1,2}|t\right\rangle =\mbox{Tr }\left\{ d\hat{\mathcal{N}}_{1,2}\,\hat{\bar{\varrho}}(t,0)\right\} $.
In these expressions the nonexact differentials are the difference
between the exact differentials 
\begin{equation}
d\mathcal{N}_{1,2}(t)=d\,\mbox{Tr }\left\{ \hat{\mathcal{N}}_{1,2}\,\hat{\bar{\varrho}}(t,0)\right\} =\sum_{\ell=1}^{s}\frac{\partial\mathcal{N}_{1,2}(t)}{\partial\lambda_{\ell}}d\,\lambda_{\ell},
\end{equation}
 and 
\begin{equation}
\left\langle d\hat{\mathcal{N}}_{1,2}|t\right\rangle =\mbox{Tr }\left\{ \frac{\partial\hat{\mathcal{N}}_{1,2}(t)}{\partial\lambda_{\ell}}d\,\lambda_{\ell}\,\hat{\bar{\varrho}}(t,0)\right\} ,
\end{equation}
 the latter being the average value of the change in the corresponding
dynamical quantity due to the modification of the control parameters.

Equation (\ref{eq:differential}) tells us that the non-equilibrium
thermodynamic variables $F_{1,2}(t)$ are integrating factors for
the nonexact differentials $\dbar\mathcal{N}_{1,2}(t)$.

Using expression (\ref{eq:ie_2_fluidos}) it is possible to study the role
of the Fröhlich contribution: changing the value of $\mathrm{F}$
(the coupling strength associated to Fröhlich contribution) in Eqs.
(\ref{eq:N1_evol-ajuste}), (\ref{eq:N2_evol-ajuste}) and (\ref{eq:banho_evol-ajuste}),
we may virtually compare the informational entropy in systems with
different Fröhlich coupling strengths.

\begin{figure}[h]
\begin{centering}
\psfrag{x}[t][c][0.85]{Scaled Time $\bar{t}=\frac{t}{\tau}$}
\psfrag{y}[b][t][0.85]{Informational Entropy $\bar{S}_{0,\lambda}(t)/\bar{S}^{\mathrm{eq}}$}
\psfrag{SF}[][]{$\frac{\bar{S}_{\mathrm{F}}(t)}{\bar{S}^{\mathrm{eq}}}$}
\psfrag{S0}[][]{$\frac{\bar{S}_{0}(t)}{\bar{S}^{\mathrm{eq}}}$}\includegraphics[width=7cm]{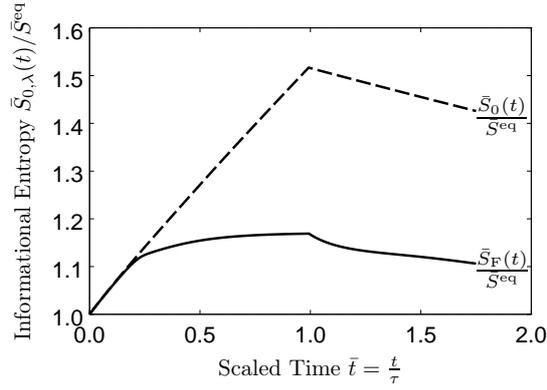}
\par\end{centering}

\caption{\label{fig:IE_time}Informational entropy as function of time associated
with magnon populations from Eqs. (\ref{eq:N1_evol-ajuste}) and (\ref{eq:N2_evol-ajuste})
displayed in Fig. \ref{fig:time}. Solid line represents the
system with Fröhlich contribution, while the dashed line refers
to a system in which the Fröhlich contribution is absent {[}$\mathrm{F}=0$
on Eqs. (\ref{eq:N1_evol-ajuste}) and (\ref{eq:N2_evol-ajuste}){]}. 
Radiation pumping switched off at $\bar{t}=1$. We recall that the values
of the different parameters are indicated in the caption of Fig. \ref{fig:time}.}
\end{figure}

The pumped system of magnons presented in Fig. \ref{fig:time} \nobreakdash-
where the magnon populations were numerically obtained from Eqs. (\ref{eq:N1_evol-ajuste})
and (\ref{eq:N2_evol-ajuste}) \nobreakdash- has the informational entropy,
obtained with the aid of Eq. (\ref{eq:ie_2_fluidos}), displayed as
function of time in Fig. \ref{fig:IE_time}. In this figure we also
show the time evolution of the informational entropy for the magnon
populations obtained from Eqs. (\ref{eq:N1_evol-ajuste}) and (\ref{eq:N2_evol-ajuste})
with $\mathrm{F}=0$, i.e., a pumped magnon system with negligible
Fröhlich contributions.

\begin{figure}[h]
\begin{centering}
\psfrag{x}[t][c][0.85]{Scaled Rate of Pumping $\mathrm{I}$}
\psfrag{y}[b][t][0.85]{Entropy $\bar{S}_{0,\mathrm{F}}(\mathrm{I})/\bar{S}^{\mathrm{eq}}$}\includegraphics[width=7cm]{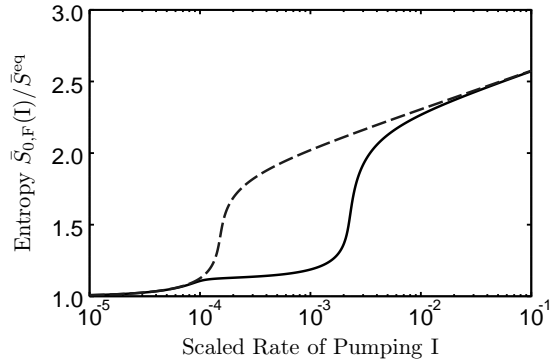}
\par\end{centering}

\caption{\label{fig:IE_ss_I}Informational entropy of the steady states as
function of the scaled rate of pumping $\mathrm{I}$ for systems with (solid) and
without (dashed) Fröhlich contribution.}
\end{figure}

It can be noticed that the informational entropy values are
lower when Fröhlich contribution is present, as it should as a result of
having increasing ordering, that is, information increase. The same behavior occur
in the case of the informational entropy of the steady states as function
of the scaled rate of pumping where, as shown in Fig. \ref{fig:IE_ss_I},
the presence of the Fröhlich contribution, precisely in the region of the 
condensate, leads to a decrease of the informational entropy.

\begin{figure}[H]
\begin{centering}
\psfrag{x}[t][c][0.85]{Scaled Rate of Pumping $\mathrm{I}$}
\psfrag{y}[b][t][0.85]{Order Parameter $\Delta(\mathrm{F},\mathrm{I})$}\includegraphics[width=7cm]{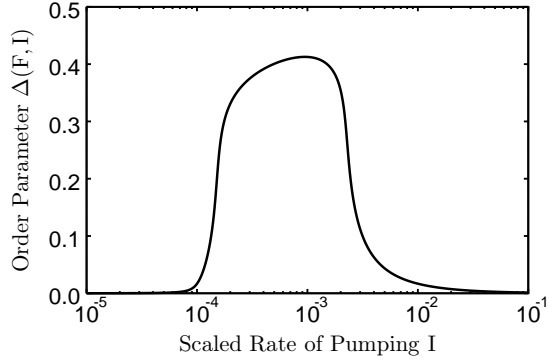}
\par\end{centering}

\caption{\label{fig:order_parameter_I}Order parameter of Eq. (\ref{eq:order_parameter})
as function of the scaled rate of pumping $\mathrm{I}$.}
\end{figure}

This decrease of informational entropy due to the Fröhlich contribution
may be understood as some kind of increase in order and, to characterize
this point, we introduce the order parameter 
\begin{equation}
\Delta(\mathrm{F},\mathrm{I})=\frac{\bar{S}_{0}^{\mathrm{S}}(\mathrm{I})-\bar{S}_{\mathrm{F}}^{\mathrm{S}}(\mathrm{I})}{\bar{S}_{0}^{\mathrm{S}}(\mathrm{I})}=1-\frac{\bar{S}_{\mathrm{F}}^{\mathrm{S}}(\mathrm{I})}{\bar{S}_{0}^{\mathrm{S}}(\mathrm{I})},\label{eq:order_parameter}
\end{equation}
 where $\bar{S}_{\mathrm{F}}^{\mathrm{S}}(\mathrm{I})$ and $\bar{S}_{0}^{\mathrm{S}}(\mathrm{I})$
are the steady-states informational entropies with and without Fröhlich contribution, that is 
\begin{align}
\bar{S}_{\mathrm{F}}^{\mathrm{S}}(\mathrm{I})=f_{1}\left\{ \left(\mathcal{N}_{1}^{\mathrm{S}}+1\right)\,\ln\left(\mathcal{N}_{1}^{\mathrm{S}}+1\right)-\mathcal{N}_{1}^{\mathrm{S}}\,\ln\left(\mathcal{N}_{1}^{\mathrm{S}}\right)\right\} +\nonumber \\
+f_{2}\left\{ \left(\mathcal{N}_{2}^{\mathrm{S}}+1\right)\,\ln\left(\mathcal{N}_{2}^{\mathrm{S}}+1\right)-\mathcal{N}_{2}^{\mathrm{S}}\,\ln\left(\mathcal{N}_{2}^{\mathrm{S}}\right)\right\} 
\end{align}
and 
\begin{align}
\bar{S}_{0}^{\mathrm{S}}(\mathrm{I})=\: & f_{1}\left\{ \left(\mathcal{N}_{1}^{\mathrm{S},\mathrm{F}=0}+1\right)\,\ln\left(\mathcal{N}_{1}^{\mathrm{S},\mathrm{F}=0}+1\right)-\mathcal{N}_{1}^{\mathrm{S},\mathrm{F}=0}\,\ln\left(\mathcal{N}_{1}^{\mathrm{S},\mathrm{F}=0}\right)\right\} +\nonumber \\
 & +f_{2}\left\{ \left(\mathcal{N}_{2}^{\mathrm{S},\mathrm{F}=0}+1\right)\,\ln\left(\mathcal{N}_{2}^{\mathrm{S},\mathrm{F}=0}+1\right)-\mathcal{N}_{2}^{\mathrm{S},\mathrm{F}=0}\,\ln\left(\mathcal{N}_{2}^{\mathrm{S},\mathrm{F}=0}\right)\right\} ,
\end{align}
 where the dependence of the steady-state populations on $\mathrm{I}$
has not been explicitly indicated. Fig. \ref{fig:order_parameter_I} presents
the order parameter as function of the scaled rate of pumping which
highlights this kind of complex order.

\begin{figure}[h]
\begin{centering}
\psfrag{x}[t][c][0.85]{Fröhlich Parameter $\mathrm{F}$}
\psfrag{ya}[b][t][0.85]{Magnon Population $\mathcal{N}_{1,2}^{\mathrm{S}}$}
\psfrag{yb}[b][t][0.85]{Order Parameter $\Delta(\mathrm{F},\mathrm{I})$}
\psfrag{a}[][]{(a)}
\psfrag{b}[][]{(b)}
\psfrag{n1}[][]{$\mathcal{N}_{1}^{\mathrm{S}}$}
\psfrag{n2}[][]{$\mathcal{N}_{2}^{\mathrm{S}}$}\includegraphics[width=0.5\columnwidth]{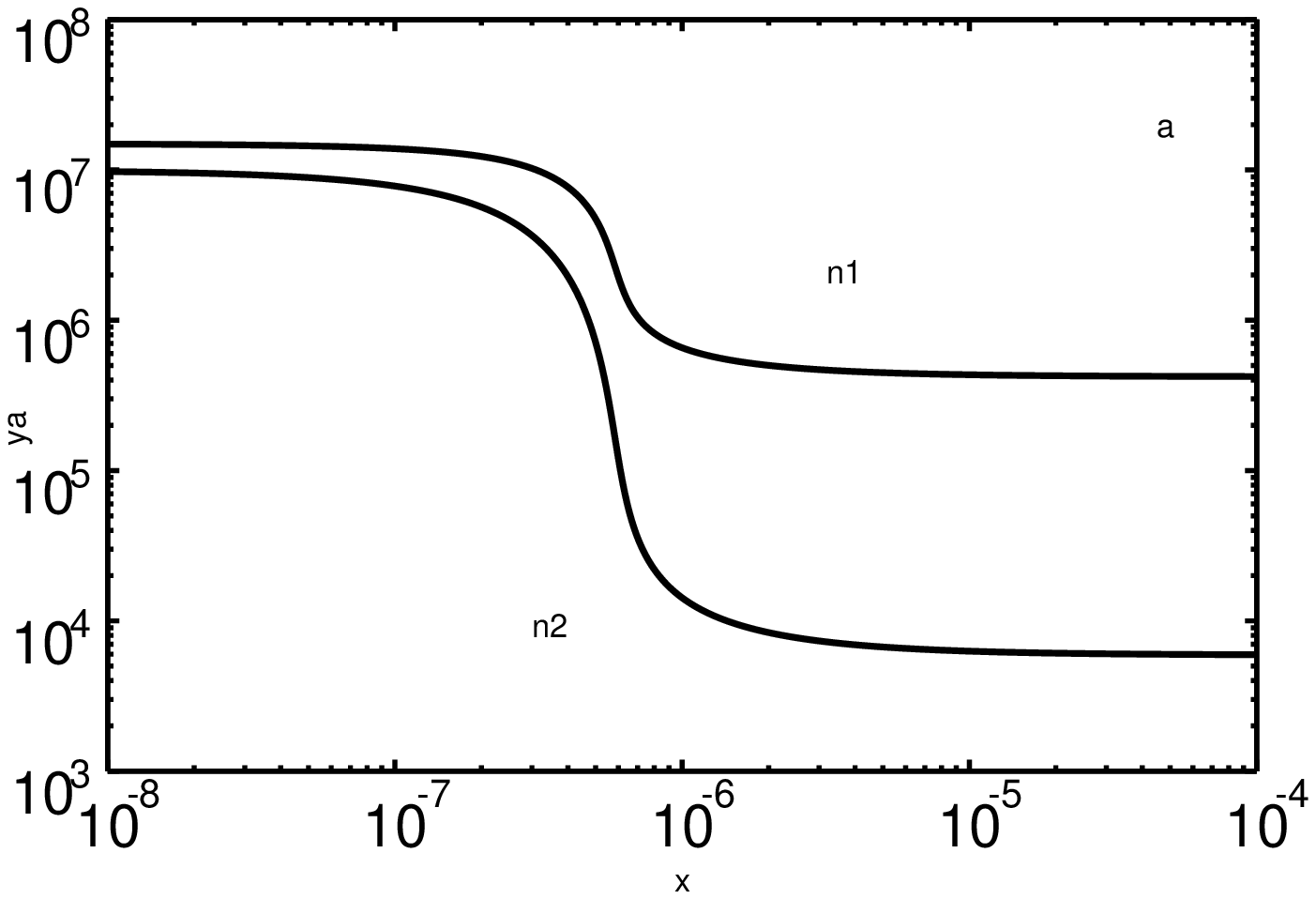}\includegraphics[width=0.5\columnwidth]{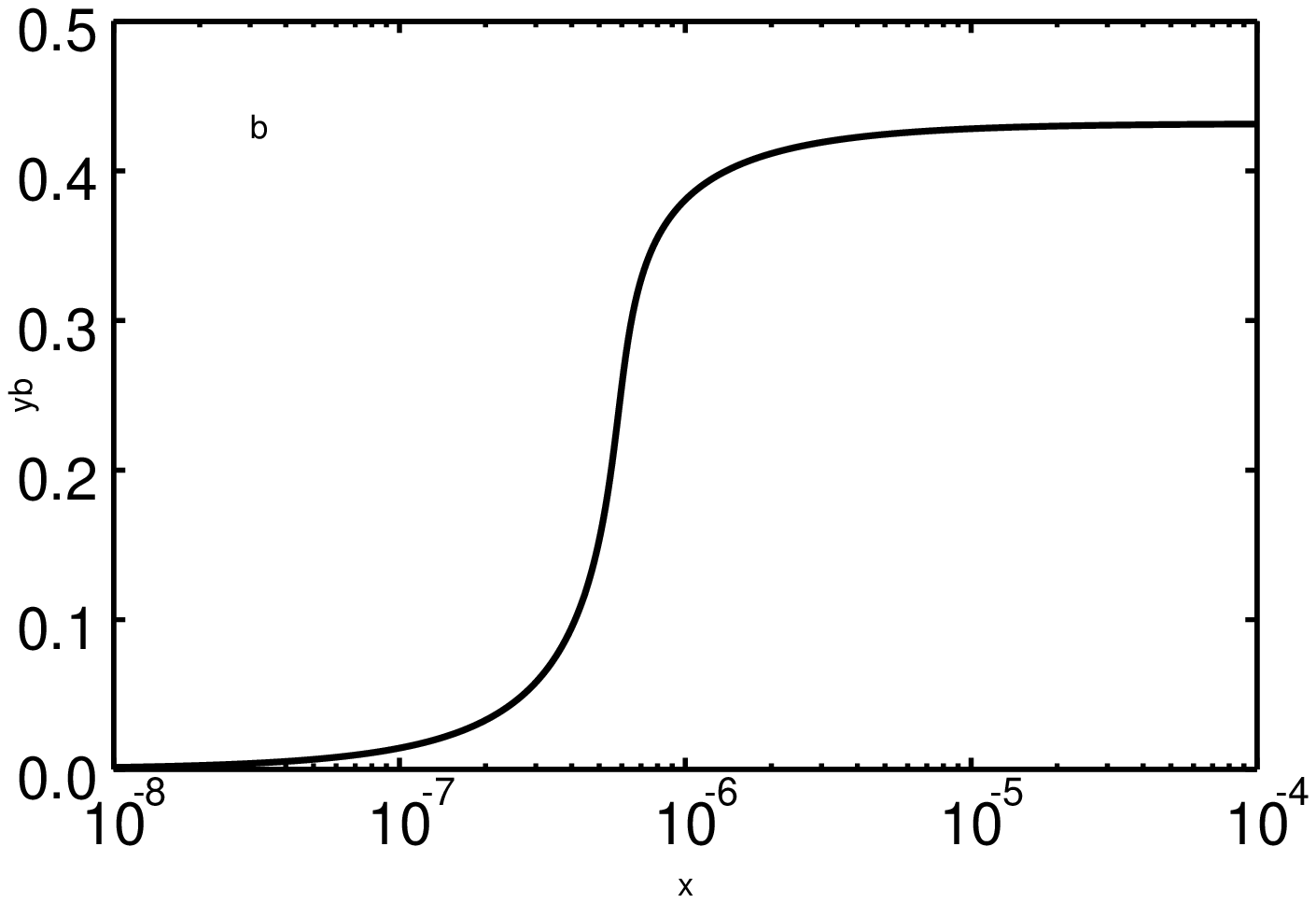}
\par\end{centering}

\caption{\label{fig:SS_F}(a) Steady-state magnon populations as function of
$\mathrm{F}$. (b) The related order parameter.}
\end{figure}

The role of the Fröhlich contribution may be evidenced through the
numerical analysis of the order parameter as function of the Fröhlich
contribution coupling strength when the rate of pumping is fixed.
In Fig. \ref{fig:SS_F} we present the mean steady-state populations
$\mathcal{N}_{1,2}^{\mathrm{S}}$ as function of $\mathrm{F}$ for
fixed scaled rate of pumping $\mathrm{I}=8\times10^{-4}$ and the
corresponding informational entropy order parameter.

As can be seen in the Fig. \ref{fig:SS_F}(a), the magnon steady-state populations
$\mathcal{N}_{1,2}^{\mathrm{S}}$ decrease as the non-linear Fröhlich
contribution coupling strength increases (notably the mean population
associated with high frequencies magnons $\mathcal{N}_{2}^{\mathrm{S}}$).
This complex behavior of the steady-state populations may be understood
considering that: (i) the Fröhlich contribution leads to the formation
of the condensate in which the magnons with lower frequency are overpopulated
at the expense of the higher in frequency populations; (ii) since
the substantial decrease of $\mathcal{N}_{2}^{\mathrm{S}}$ relative
to $\mathcal{N}_{1}^{\mathrm{S}}$ diminishes the absorbance of the
material [because of the positive feedback effect of the parallel
pumping, see Eq. (\ref{eq:N2_evol_fonte_ajuste})] the net flux of absorbed
energy is lower than in the case without Fröhlich interaction, justifying
the global fall of the mean population. The order parameter behavior,
shown in Fig. \ref{fig:SS_F}(b), corroborates the idea that Fröhlich
contribution enhances the complex order mentioned before.

\subsection{IST entropy production}
\label{sec:entropy_prod}

We analyze the informational-entropy production and, using Eq.
(\ref{eq:ie_bosons_1}) {[}paying attention to the logarithm of the
partition functions in Eq. (\ref{generating_functional}){]}, it can
be shown that it is given by 
\begin{equation}
\bar{\sigma}(t)=\frac{d}{dt}\bar{S}(t)=\beta_{0}\frac{dE_{\mathrm{B}}(t)}{dt}+\sum_{\mathbf{q}}F_{\mathbf{q}}(t)\frac{d\mathcal{N}_{\mathbf{q}}(t)}{dt},\label{eq:entropy_production_time}
\end{equation}
and then, taking into account Eqs. (\ref{eq:N_q_evol}) and (\ref{eq:banho}), 
it can be rewritten in terms of two contributions 
\begin{equation}
\bar{\sigma}(t)=\bar{\sigma}_{\mathrm{i}}(t)+\bar{\sigma}_{\mathrm{e}}(t),
\end{equation}
 consisting of the so-called internal one, $\bar{\sigma}_{\mathrm{i}}(t)$,
which results from internal interactions in the system, and the external
one, $\bar{\sigma}_{\mathrm{e}}(t)$, related to the interactions
with the surroundings, in this case with the source and the thermal
reservoir. They are given by 
\begin{align}
\bar{\sigma}_{\mathrm{i}}(t)= & \sum_{\mathbf{q}} \{ F_{\mathbf{q}}(t)\,\left[\mathfrak{R}_{\mathbf{q}}(t)+L_{\mathbf{q}}(t)+\mathfrak{F}_{\mathbf{q}}(t)+\mathfrak{M}_{\mathbf{q}}(t)\right]- \nonumber \\
& -\beta_{0}\hbar\omega_{\mathbf{q}}\left[\mathfrak{R}_{\mathbf{q}}(t)+L_{\mathbf{q}}(t)+\mathfrak{F}_{\mathbf{q}}(t)\right] \}, 
\end{align}
\begin{equation}
\bar{\sigma}_{\mathrm{e}}(t)=\sum_{\mathbf{q}}\left\{ F_{\mathbf{q}}(t)\,\mathfrak{S}_{\mathbf{q}}(t)+\beta_{0}\, J_{\mathrm{TD}}^{(2)}(t)\right\} ,
\end{equation}
or, using Eq. (\ref{eq:banho_estacionario}), 
\begin{equation}
\bar{\sigma}_{\mathrm{e}}(t)=\sum_{\mathbf{q}}\left\{ F_{\mathbf{q}}(t)\,\mathfrak{S}_{\mathbf{q}}(t)+\beta_{0}\hbar\omega_{\mathbf{q}}\left[\mathfrak{R}_{\mathbf{q}}(t)+L_{\mathbf{q}}(t)+\mathfrak{F}_{\mathbf{q}}(t)\right]\right\} .
\end{equation}

In the two-fluid model the informational-entropy production is thus given by
\begin{align}
 & \bar{\sigma}_{\mathrm{i}}(t)= \nonumber\\
= & \sum_{\mathbf{q}}\left\{ \left[F_{\mathbf{q}}(t)-\beta_{0}\hbar\omega_{\mathbf{q}}\right]\,\left[\mathfrak{R}_{\mathbf{q}}(t)+L_{\mathbf{q}}(t)+\mathfrak{F}_{\mathbf{q}}(t)\right]+F_{\mathbf{q}}(t)\,\mathfrak{M}_{\mathbf{q}}(t)\right\} \approx\\
\approx & \left\{ \ln\left(\frac{\mathcal{N}_{1}+1}{\mathcal{N}_{1}}\right)-\beta_{0}\hbar\omega_{1}\right\} \,\sum_{\mathbf{q}\in R_{1}}\left[\mathfrak{R}_{\mathbf{q}}(t)+L_{\mathbf{q}}(t)+\mathfrak{F}_{\mathbf{q}}(t)\right]+ \nonumber\\
 & +\left\{ \ln\left(\frac{\mathcal{N}_{2}+1}{\mathcal{N}_{2}}\right)-\beta_{0}\hbar\omega_{2}\right\} \,\sum_{\mathbf{q}\in R_{2}}\left[\mathfrak{R}_{\mathbf{q}}(t)+L_{\mathbf{q}}(t)+\mathfrak{F}_{\mathbf{q}}(t)\right]+ \nonumber\\
 & +\ln\left(\frac{\mathcal{N}_{1}+1}{\mathcal{N}_{1}}\right)\,\sum_{\mathbf{q}\in R_{1}}\mathfrak{M}_{\mathbf{q}}(t)+\ln\left(\frac{\mathcal{N}_{2}+1}{\mathcal{N}_{2}}\right)\,\sum_{\mathbf{q}\in R_{2}}\mathfrak{M}_{\mathbf{q}}(t)= \nonumber\\
= & -\frac{n}{\tau}\left\{ \ln\left(\frac{\mathcal{N}_{1}+1}{\mathcal{N}_{1}}\right)-\beta_{0}\hbar\omega_{1}\right\} \left\{ \mathrm{D}\,\mathcal{N}_{1}(\mathcal{N}_{1}-\mathcal{N}_{1}^{(0)})+f_{1}\left[\mathcal{N}_{1}-\mathcal{N}_{1}^{(0)}\right]\right\} + \nonumber\\
 & +\frac{n}{\tau}\left\{ \ln\left(\frac{\mathcal{N}_{1}+1}{\mathcal{N}_{1}}\right)-\beta_{0}\hbar\omega_{1}\right\} \,\mathrm{F}\left\{ \mathcal{N}_{1}\mathcal{N}_{2}+\left(\bar{\nu}+1\right)\mathcal{N}_{2}-\bar{\nu}\mathcal{N}_{1}\right\} - \nonumber\\
 & -\frac{n}{\tau}\left\{ \ln\left(\frac{\mathcal{N}_{2}+1}{\mathcal{N}_{2}}\right)-\beta_{0}\hbar\omega_{2}\right\} \left\{ \mathrm{D}\,\mathcal{N}_{2}(\mathcal{N}_{2}-\mathcal{N}_{2}^{(0)})+f_{2}\left[\mathcal{N}_{2}-\mathcal{N}_{2}^{(0)}\right]\right\} - \nonumber\\
 & -\frac{n}{\tau}\left\{ \ln\left(\frac{\mathcal{N}_{2}+1}{\mathcal{N}_{2}}\right)-\beta_{0}\hbar\omega_{2}\right\} \,\mathrm{F}\left\{ \mathcal{N}_{1}\mathcal{N}_{2}+\left(\bar{\nu}+1\right)\mathcal{N}_{2}-\bar{\nu}\mathcal{N}_{1}\right\} + \nonumber\\
 & +\frac{n}{\tau}\left\{ \ln\left(\frac{\mathcal{N}_{2}+1}{\mathcal{N}_{2}}\right)-\ln\left(\frac{\mathcal{N}_{1}+1}{\mathcal{N}_{1}}\right)\right\} \mathrm{M}\left\{ \mathcal{N}_{1}\left(\mathcal{N}_{1}+1\right)+\mathcal{N}_{2}\left(\mathcal{N}_{2}+1\right)\right\}\times \nonumber \\
& \qquad \times (\mathcal{N}_{1}\frac{\mathcal{N}_{2}^{(0)}}{\mathcal{N}_{1}^{(0)}}-\mathcal{N}_{2}),
\end{align}
 and 
\begin{align}
 & \bar{\sigma}_{\mathrm{e}}(t)=\nonumber \\
= & \sum_{\mathbf{q}}\left\{ F_{\mathbf{q}}(t)\,\mathfrak{S}_{\mathbf{q}}(t)+\beta_{0}\hbar\omega_{\mathbf{q}}\left[\mathfrak{R}_{\mathbf{q}}(t)+L_{\mathbf{q}}(t)+\mathfrak{F}_{\mathbf{q}}(t)\right]\right\} \approx\nonumber \\
\approx & \ln\left(\frac{\mathcal{N}_{1}+1}{\mathcal{N}_{1}}\right)\,\sum_{\mathbf{q}\in R_{1}}\mathfrak{S}_{\mathbf{q}}(t)+\ln\left(\frac{\mathcal{N}_{2}+1}{\mathcal{N}_{2}}\right)\,\sum_{\mathbf{q}\in R_{2}}\mathfrak{S}_{\mathbf{q}}(t)+\nonumber \\
 & +\beta_{0}\hbar\omega_{1}\sum_{\mathbf{q}\in R_{1}}\left[\mathfrak{R}_{\mathbf{q}}(t)+L_{\mathbf{q}}(t)+\mathfrak{F}_{\mathbf{q}}(t)\right]+ \nonumber\\
& +\beta_{0}\hbar\omega_{2}\sum_{\mathbf{q}\in R_{2}}\left[\mathfrak{R}_{\mathbf{q}}(t)+L_{\mathbf{q}}(t)+\mathfrak{F}_{\mathbf{q}}(t)\right]=\nonumber \\
= & \frac{n}{\tau}\ln\left(\frac{\mathcal{N}_{2}+1}{\mathcal{N}_{2}}\right)\,\mathrm{I}\,(1+2\mathcal{N}_{2})+\nonumber \\
 & +\frac{n}{\tau}\,\beta_{0}\hbar\omega_{1}\left\{ -\mathrm{D}\,\mathcal{N}_{1}(\mathcal{N}_{1}-\mathcal{N}_{1}^{(0)})-f_{1}\left[\mathcal{N}_{1}-\mathcal{N}_{1}^{(0)}\right]\right.+\nonumber \\
& \qquad\qquad\qquad + \mathrm{F}\left[ \mathcal{N}_{1}\mathcal{N}_{2}+\left(\bar{\nu}+1\right)\mathcal{N}_{2}-\bar{\nu}\mathcal{N}_{1}\right] \Big\} +\nonumber \\
 & +\frac{n}{\tau}\,\beta_{0}\hbar\omega_{2}\left\{ -\mathrm{D}\,\mathcal{N}_{2}(\mathcal{N}_{2}-\mathcal{N}_{2}^{(0)})-f_{2}\left[\mathcal{N}_{2}-\mathcal{N}_{2}^{(0)}\right] \right. -\nonumber \\
& \qquad\qquad\qquad - \mathrm{F}\left[ \mathcal{N}_{1}\mathcal{N}_{2}+\left(\bar{\nu}+1\right)\mathcal{N}_{2}-\bar{\nu}\mathcal{N}_{1}\right] \Big\} ,
\end{align}
 where we used that 
\begin{equation}
\ln\left(\frac{\mathcal{N}_{1,2}^{(0)}+1}{\mathcal{N}_{1,2}^{(0)}}\right)=\beta_{0}\hbar\omega_{1,2},
\end{equation}
with $\mathcal{N}_{1,2}^{(0)}$ being the distribution in equilibrium.

\begin{figure}[H]
\begin{centering}
\psfrag{x}[t][c][0.85]{Scaled Time $\bar{t}=t/\tau$}
\psfrag{y}[b][t][0.85]{Entropy Production $\sigma(\bar{t})$}
\psfrag{si}[][]{$\sigma_\mathrm{i}$}
\psfrag{se}[][]{$\sigma_\mathrm{e}$}
\psfrag{s}[][]{$\sigma$}\includegraphics[width=0.5\columnwidth]{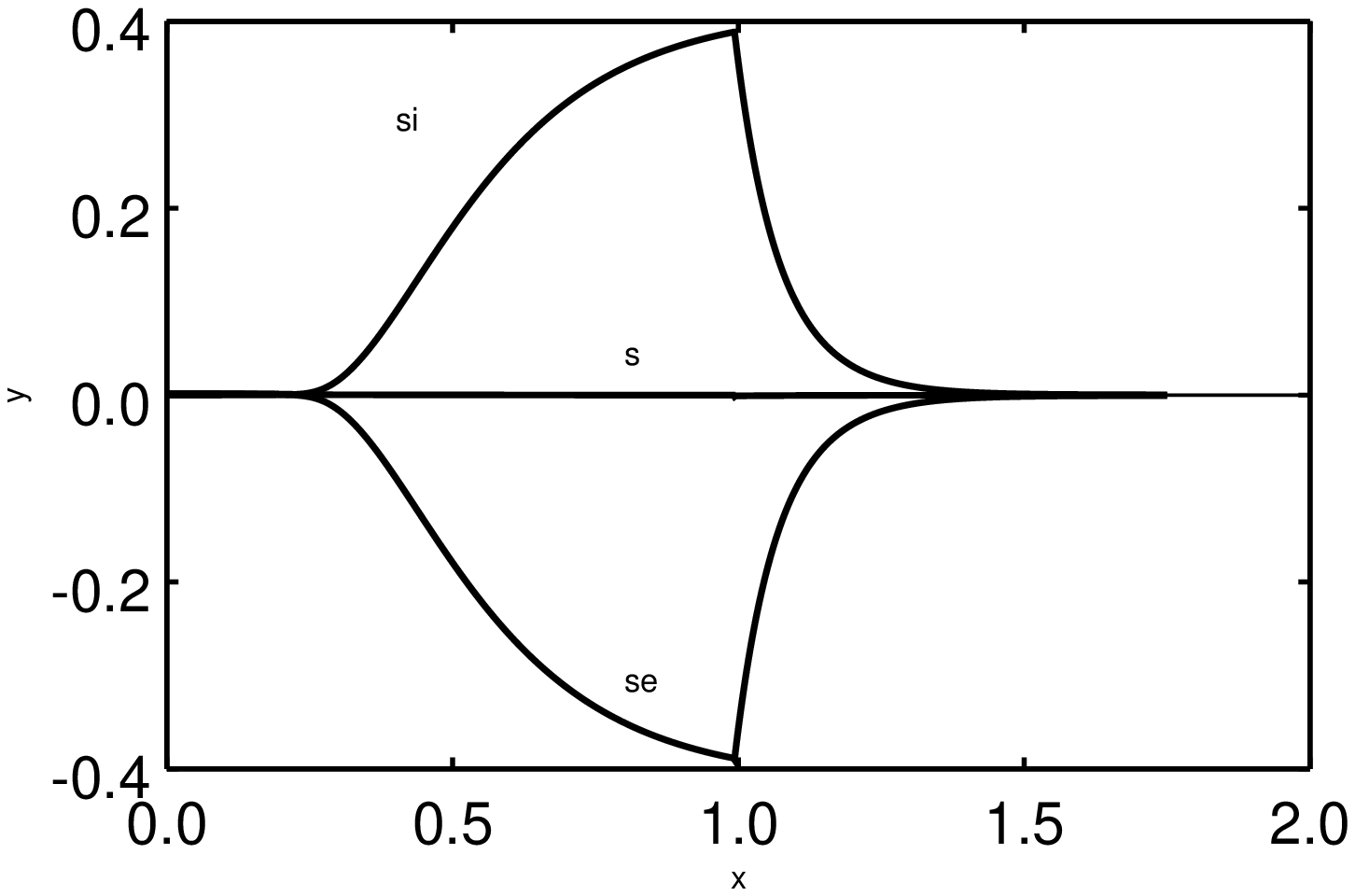}\includegraphics[width=0.5\columnwidth]{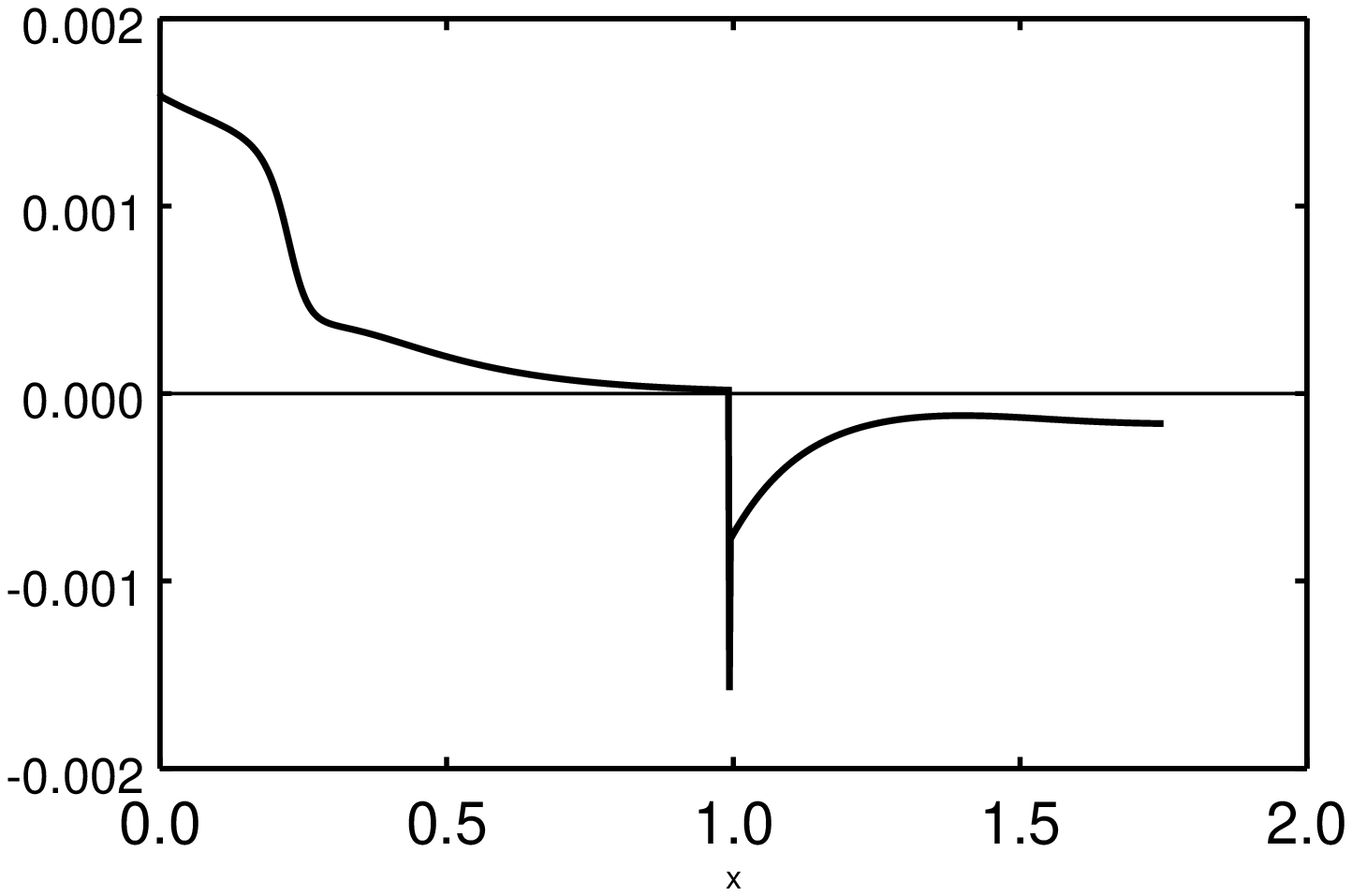}
\par\end{centering}

\caption{\label{fig:entropy_production_time}Informational-entropy production
of the system of magnons as function of time (associated with Figs.
\ref{fig:time} and \ref{fig:IE_time}). On the left it can be observed
the internal, external and total entropy production (and we call attention
to the expected non-negative values of the internal entropy production);
On the right only the total entropy production is shown. We recall that the pumping
source is switched off at $\bar{t}=1$.}
\end{figure}

In Fig. \ref{fig:entropy_production_time} we show the informational-entropy
production of the system of magnons evolving in time (i.e., the entropy
production associated with the evolution described in Fig. \ref{fig:time}).
We can observe that, although the internal entropy production is strictly non-negative \nobreakdash- as it should \nobreakdash-, the total entropy 
production may have positive and negative values. 
This is related, in informational non-equilibrium statistical thermodynamics, with the generalized $\mathscr{H}$-theorem in the sense of Jancel
\cite{jancel1969}, which we have called \emph{weak principle of increasing
of informational entropy}, namely, given the informational statistical
entropy $\bar{S}(t)$ of Eq. (\ref{eq:ie_bosons_1}) and the informational entropy production $\sigma(t)$ of Eq. (\ref{eq:entropy_production_time}),
the principle tells us that 
\begin{align}
\Delta\bar{S}(t)=\: & \bar{S}(t)-\bar{S}(t_{0})=\nonumber \\
=\: & \int_{-\infty}^{t}dt'\,\sigma(t')\geq0.\label{eq:teoremaH}
\end{align}

Equation (\ref{eq:teoremaH}) does not prove that $\sigma(t)$ is
a monotonically increasing function of time, as required by phenomenological
irreversible thermodynamic theories. We have only proved the weak
condition that as the system evolves it is predominantly definite positive.
We stress the fact that this result is a consequence of the presence
of the irreversible part of $\hat{\rho}_{\varepsilon}(t)$ not contained
in $\hat{\bar{\rho}}(t)$, which is then, as stated previously, the
part that accounts for - in the description of the macroscopic state
of the system - the processes which generate dissipation. Furthermore,
the informational entropy with the evolution property of Eq. (\ref{eq:teoremaH})
is the coarse-grained entropy of Eq. (\ref{eq:entropy_formal}), the
coarse-graining being performed by the action of the projection operator
$\mathcal{P}_{\varepsilon}(t)$ of Eqs. (\ref{eq:projetor1}) and
(\ref{eq:projetor2}): This projection operation extracts from the
Gibbs entropy the contribution associated to the constraints {[}cf.
set (\ref{eq:variaveis_termodinamicas}){]} imposed on the system,
by projecting it onto the subspace spanned by the basic dynamical
quantities (see also \cite{balian1986}). Hence, the informational
entropy thus defined depends on the choice of the basic set of macroscopic
variables, whose completeness in a purely thermodynamic sense cannot
be indubitably asserted. We restate that in each particular problem
under consideration the information lost as a result of the particular
truncation of the set of basic variables must be carefully evaluated
\cite{vasconcellos1991,ramos2000}. Retaking the question of the signal
of $\sigma(t)$, we conjecture that it is always non-negative, since
it can not be intuitively understood how information can be gained
in some time intervals along the irreversible evolution of the system.
However, this is expected to be valid as long as we are using an,
in a sense, complete description of the system. Once a truncation
procedure is introduced \cite{ramos2000} the local density of informational
entropy production is no longer monotonously increasing in time; this
has been illustrated by Criado-Sancho and Llebot \cite{criado-sancho1993}
in the realm of Extended Irreversible Thermodynamics, and in IST in
\cite{ramos1998}. The reason is, as pointed out by Balian \emph{et
al.}\cite{balian1986} that the truncation procedure introduces some
kind of additional (spurious) information at the step when the said
truncation is imposed.

\begin{figure}[H]
\begin{centering}
\psfrag{x}[t][c][0.85]{Intensity $\mathrm{I}$}
\psfrag{y}[b][t][0.85]{Entropy Production $\sigma(\bar{t})$}
\psfrag{s}[][]{$\sigma_\mathrm{i}=|\sigma_\mathrm{e}|$}{$\sigma$}\includegraphics[width=7cm]{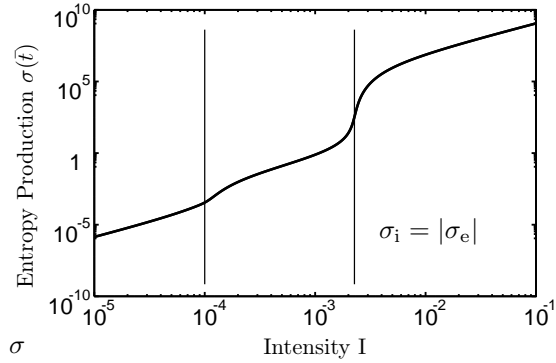}
\par\end{centering}

\caption{\label{fig:SS_entropy_production} Internal and external informational-entropy
production absolute values of the system in steady state as function of the scaled rate of pumping $\mathrm{I}$ (related
to Fig. \ref{fig:stationary}); $\sigma_\mathrm{i}=-\sigma_\mathrm{e}$}
\end{figure}

Another important result, shown in Fig. \ref{fig:SS_entropy_production},
is the informational-entropy production for the steady states (see
Fig. \ref{fig:stationary}).

\subsection{The Evolution Criterion}
\label{sec:evol_crit}

The change in time of IST-entropy production can be separated into
two parts, namely

\begin{equation}
\frac{d}{dt}\bar{\sigma}(t)=\frac{d_{Q}}{dt}\bar{\sigma}(t)+\frac{d_{F}}{dt}\bar{\sigma}(t),
\end{equation}
 where 
\begin{equation}
\frac{d_{Q}}{dt}\bar{\sigma}(t)=\sum_{\mathbf{q}}F_{\mathbf{q}}(t)\frac{d^{2}\mathcal{N}_{\mathbf{q}}(t)}{dt^{2}},
\end{equation}
 that is, the part that accounts for the change in time of $\mathcal{N}_{\mathbf{q}}(t)$, and 
\begin{equation}
\frac{d_{F}}{dt}\bar{\sigma}(t)=\sum_{\mathbf{q}}\frac{dF_{\mathbf{q}}(t)}{dt}\:\frac{d\mathcal{N}_{\mathbf{q}}(t)}{dt},\label{eq:forceXflux}
\end{equation}
accounting for the part of change in time of the non-equilibrium thermodynamics variables $F_{\mathbf{q}}(t)$.
Recalling that $F_{\mathbf{q}}(t)$ may be expressed in terms of the populations {[}cf. Eq. (\ref{eq:F(populacao)-1}){]},
we have that 
\begin{align}
\frac{dF_{\mathbf{q}}(t)}{dt}= & \frac{d}{dt}\ln\left\{ \frac{\mathcal{N}_{\mathbf{q}}(t)+1}{\mathcal{N}_{\mathbf{q}}(t)}\right\} = \nonumber \\
= & \left[\frac{1}{\mathcal{N}_{\mathbf{q}}(t)+1}-\frac{1}{\mathcal{N}_{\mathbf{q}}(t)}\right]\frac{d\mathcal{N}_{\mathbf{q}}(t)}{dt}=-\frac{1}{\left(\mathcal{N}_{\mathbf{q}}+1\right)\mathcal{N}_{\mathbf{q}}}\,\frac{d\mathcal{N}_{\mathbf{q}}(t)}{dt},\label{eq:dF/dt}
\end{align}
 and therefore
\begin{equation}
\frac{d_{F}}{dt}\bar{\sigma}(t)=-\sum_{\mathbf{q}}\frac{1}{\left(\mathcal{N}_{\mathbf{q}}+1\right)\mathcal{N}_{\mathbf{q}}}\,\left(\frac{d\mathcal{N}_{\mathbf{q}}}{dt}\right)^{2}\leq0.\label{eq:GP}
\end{equation}

This inequality verifies for this system the generalization of Glansdorff\textendash{}Prigogine's
thermodynamic criterion of evolution\cite{luzzi2000,luzzi2001,glansdorff1971}. That is, along the trajectory
of the macrostate of the system in the thermodynamic (or Gibbs) space
of states, the quantity of Eq. (\ref{eq:forceXflux}) is always non-positive,
a quantity which in classical Onsagerian thermodynamics is the product
of the change in time of the thermodynamic forces times the fluxes
of matter and energy.

\subsection{The (In)stability Criterion}
\label{sec:stab_crit}

Within the above discussed framework of a non-equilibrium thermodynamics
of the Fröhlich-Bose-Einstein condensation of magnons, we may analyze
again the stability of the steady-states populations $\mathcal{N}_{\mathbf{q}}^{\mathrm{S}}$.
Considering arbitrary small deviations, say $\epsilon\,\eta_{\mathbf{q}}(t)$, from the steady state, we may expand
the informational entropy in the form

\begin{equation}
\bar{S}\left(\left\{ \mathcal{N}_{\mathbf{q}}(t)\right\} \right)=\bar{S}\left(\left\{ \mathcal{N}_{\mathbf{q}}^{\mathrm{S}}+\epsilon\eta_{\mathbf{q}}(t)\right\} \right)=\bar{S}\left(\left\{ \mathcal{N}_{\mathbf{q}}^{\mathrm{S}}\right\} \right)+\delta\bar{S}+\delta^{2}\bar{S}+\dots\:,
\end{equation}
 with 
\begin{equation}
\delta^{\mathrm{n}}\bar{S}=\left.\frac{\partial^{\mathrm{n}}\bar{S}}{\partial\epsilon^{\mathrm{n}}}\right|_{\epsilon=0}\frac{\epsilon^{\mathrm{n}}}{\mathrm{n}!}.
\end{equation}

Since 
\begin{equation}
\frac{\partial^{2}\bar{S}}{\partial\epsilon^{2}}=-{\displaystyle \sum_{\mathbf{q}}}\frac{\eta_{\mathbf{q}}^{2}(t)}{\left[\mathcal{N}_{\mathbf{q}}^{\mathrm{S}}+\epsilon\eta_{\mathbf{q}}(t)+1\right]\left[\mathcal{N}_{\mathbf{q}}^{\mathrm{S}}+\epsilon\eta_{\mathbf{q}}(t)\right]},
\end{equation}

we have that the second variation of the entropy is 
\begin{equation}
\delta^{2}\bar{S}=-{\displaystyle \sum_{\mathbf{q}}}\frac{\epsilon^{2}\eta_{\mathbf{q}}^{2}(t)}{\left(\mathcal{N}_{\mathbf{q}}^{\mathrm{S}}+1\right)\mathcal{N}_{\mathbf{q}}^{\mathrm{S}}}=-{\displaystyle \sum_{\mathbf{q}}}\frac{\left[\Delta\mathcal{N}_{\mathbf{q}}(t)\right]^{2}}{\left(\mathcal{N}_{\mathbf{q}}^{\mathrm{S}}+1\right)\mathcal{N}_{\mathbf{q}}^{\mathrm{S}}}\leq0,\label{eq:second_variation}
\end{equation}
 where $\Delta\mathcal{N}_{\mathbf{q}}(t)$ represents the value of
the imposed arbitrary deviation from the steady state and the non-positiveness
of Eq. (\ref{eq:second_variation}) is a manifestation of the convexity
of the maximized informational entropy. Differentiation in time of
Eq. (\ref{eq:second_variation}) introduces the quantity called \emph{excess
of entropy production function}, namely 
\begin{equation}
\delta^{2}\bar{\sigma}(t)=\frac{1}{2}\,\frac{d}{dt}\delta^{2}\bar{S}(t)=-{\displaystyle \sum_{\mathbf{q}}}\frac{\Delta\mathcal{N}_{\mathbf{q}}(t)}{\left(\mathcal{N}_{\mathbf{q}}^{\mathrm{S}}+1\right)\mathcal{N}_{\mathbf{q}}^{\mathrm{S}}}\,\frac{d}{dt}\Delta\mathcal{N}_{\mathbf{q}}(t),
\end{equation}
 which, in the two fluid model has the following form 
\begin{equation}
\delta^{2}\bar{\sigma}(\mathrm{I},t)=-\frac{n_{1}\,\Delta\mathcal{N}_{1}(\mathrm{I},t)}{\left[\mathcal{N}_{1}^{\mathrm{S}}(\mathrm{I})+1\right]\mathcal{N}_{1}^{\mathrm{S}}(\mathrm{I})}\,\frac{d}{dt}\Delta\mathcal{N}_{1}(\mathrm{I},t)-\frac{n_{2}\,\Delta\mathcal{N}_{2}(\mathrm{I},t)}{\left[\mathcal{N}_{2}^{\mathrm{S}}(\mathrm{I})+1\right]\mathcal{N}_{2}^{\mathrm{S}}(\mathrm{I})}\,\frac{d}{dt}\Delta\mathcal{N}_{2}(\mathrm{I},t).
\end{equation}

According to Glansdorff-Prigogine (in)stability criterion\cite{luzzi2000,luzzi2001,glansdorff1971}, if

\begin{equation}
\frac{1}{2}\,\delta^{2}\bar{S}(t)\,\delta^{2}\bar{\sigma}(t)\leq0 \label{eq:gp_criterion}
\end{equation}
 then the steady-state is stable, and this is so once
$\displaystyle{\Delta\mathcal{N}_{1,2}\frac{d\mathcal{N}_{1,2}}{dt}\leq 0}$, as it follows
from solving the evolution equations, Eqs. (\ref{eq:N1_evol-ajuste}) and 
(\ref{eq:N2_evol-ajuste}). Then $\delta^{2}\bar{\sigma}\geq 0$ and, since
$\delta^{2}\bar{S}\leq0$, Eq. (\ref{eq:gp_criterion}) is verified. Hence, 
for the given constraints the reference steady state is stable for all 
fluctuations compatible with the equations of evolution.

The stability has been derived in relation to homogeneous fluctuations, but 
instability may emerge due to induced inhomogeneous states. Some experimental
observations may be pointing in that direction, and the question is under consideration.

Using linear stability analysis it was previously shown the stability 
of the resulting steady state (cf. Fig. \ref{fig:Lyapunov-Exponents}). Such 
stability has been here rederived using Glansdorff-Prigogine analysis which 
involves physical considerations instead of only mathematical ones, with
Glandsdorff-Prigogine's excess entropy production function being a 
Lyapunov function for this system.

\section{Concluding Remarks}

We have considered the non-equilibrium statistical thermodynamics
of the Bose-Einstein condensation of magnons excited under the action
of radio-frequency radiation pumping. It has better referred to as
Fröhlich-Bose-Einstein condensation, once, as noticed, the phenomenon which is
possible to emerge in systems of bosons, was originally evidenced
by Herbert Fröhlich\cite{froehlich1970,froehlich1980}. 

It constitutes an example of complexity in which, after a certain
threshold in the value of the intensity of the pumping source has
been attained, there follows that the energy pumped on the system
is transfered from modes higher in frequency to those lower in frequency
in a cascade-down-type process. The modes of lowest energy are then
largely populated at the expense of the other modes with higher frequencies.
The emergence of the phenomenon is driven by a nonlinear interaction
(involving two magnons and one phonon), which is in competition (kind
of ``tug of war'') with the magnon-magnon interaction. As a result,
it can be evidenced the existence of three regimes depending on the
intensity of the pumping source: at low intensities there follows
a simple linear behavior, as it should according to Prigogine's principle
of minimum local-equilibrium entropy; followed, as the pumping intensity
is increased, by a regime of formation of the Fröhlich-Bose-Einstein
condensate; and finally at a further threshold of pumping intensity
there follows a regime of simple thermal distribution (when the magnon-magnon
interaction overcomes the non-linear interaction that drives the emergence
of the condensate).

We have introduced a ``two-fluid model'' and the coefficients present
in the kinetic equations were evaluated and finally adjusted using
comparison with the experimental results in YIG. 

In the non-equilibrium thermodynamic analysis we have considered several 
important characteristics. First we have derived the so-called informational 
entropy for the system of magnons (Section \ref{sec:IST_entropy}). In Section
\ref{sec:fluctuations} are considered the fluctuations of the population operator
and the associated Maxwell relations. Furthermore, we have shown the derivation 
of a generalized $\mathscr{H}$-theorem, with the derivation of a Boltzmann-like 
relation for the non-equilibrium statistical entropy, that is, given by the 
logarithm of the number of complexions compatible with the non-equilibrium 
macroscopic constraints imposed on the system (Section \ref{sec:boltzmann}).

In Section \ref{sec:order_parameter} we have specified the magnon informational 
entropy for the ``two-fluid model'' and shown that the system obeys a kind 
of generalized Clausius relations. Then, in terms of entropy production, we 
introduced an order parameter. Through this order parameter, the informational 
entropy is shown to be smaller when the nonlinear interaction responsible 
for the onset of the NEFBEC predominates, thus evidencing the increase of 
order due to the Fröhlich interaction.

It has also been calculated the informational-entropy production function 
(Section \ref{sec:entropy_prod}), characterizing the contributions of the 
internal and external informational entropy production, with the former 
having, as it should, non-negative values characterizing dissipation, while 
the external one is negative as a result of the pumping on the system. 

Finally, it has been verified that Glansdorff-Prigogine evolution
criterion is satisfied (Section \ref{sec:evol_crit}), and from the generalization 
of Glansdorff-Prigogine (in)stability principle we have shown that the 
non-equilibrium thermodynamic state of system is stable under any condition 
(Section \ref{sec:stab_crit}). We stress that such instability has been derived 
in relation to other possible homogeneous state, but instability against 
the onset of a spatially ordered state cannot be ruled out, and is being 
under consideration.

%

\end{document}